\documentclass{article}

\usepackage[numbers,sort&compress]{natbib}
\usepackage[eandd, final]{neurips_2026}

\usepackage[utf8]{inputenc} 
\usepackage[T1]{fontenc}    
\usepackage{hyperref}       
\usepackage{url}            
\usepackage{booktabs}       
\usepackage{colortbl}
\usepackage{amsfonts}       
\usepackage{nicefrac}       
\usepackage{microtype}      
\usepackage{xcolor}         
\usepackage[table]{xcolor}
\usepackage{enumitem}
\usepackage{arydshln}
\usepackage{array}
\usepackage{makecell}
\usepackage{graphicx}
\usepackage{subcaption}
\usepackage{multirow}
\usepackage{soul}
\usepackage{fontawesome5}
\usepackage{afterpage}

\usepackage{pifont}
\newcommand{\cmark}{\textcolor{green!60!black}{\ding{51}}}
\newcommand{\xmark}{\textcolor{red!70!black}{\ding{55}}}

\newcommand{\revmode}{0}
\ifnum\revmode=0   
\fi
\ifnum\revmode=1   

\fi
\ifnum\revmode=2   

\fi

\usepackage{tcolorbox}

\definecolor{boxbluecolor}{HTML}{1F7FC1}
\newcommand{\boxblue}[1]{{\setlength{\fboxsep}{3pt}\colorbox{boxbluecolor!70}{#1}}}

\tcbuselibrary{skins}
\usepackage{wrapfig}
\usepackage{subcaption}
\usepackage{caption}

\newcommand{\promptbox}[2]{
  \begin{tcolorbox}[
    enhanced,
    colback=gray!10,
    colframe=black,
    arc=8pt,
    boxrule=0.8pt,
    drop shadow={black!30},
    left=8pt, right=8pt, top=6pt, bottom=6pt,
    title={\textbf{Prompt #1}},
    colbacktitle=black,
    coltitle=white
  ]
  #2
  \end{tcolorbox}
}

\newcommand{\Fref}[1]{Figure~\ref{#1}}
\newcommand{\Sref}[1]{Section~\ref{#1}}
\newcommand{\Tref}[1]{Table~\ref{#1}}

\newcommand{\ours}{AVENUE} 
\newcommand{\aved}{AvED} 
\newcommand{\cave}{CAVE} 
\newcommand{\cat}{Cat.}

\newcommand{\supple}{App.} 

\definecolor{Audiocolor}{HTML}{E84848}
\definecolor{Videocolor}{HTML}{F07820}
\definecolor{AVcolor}{HTML}{3CB96B}

\title{\faRocket\ AVENUE: \textbf{\underline{A}udio-\underline{V}ideo \underline{E}diti\underline{N}g \\ \underline{U}nderstanding and \underline{E}valuation}}

\author{%
  Hayeon Kim\textsuperscript{1,2}\thanks{Equal contribution.} \quad
  Yoojin Jang\textsuperscript{1}\footnotemark[1] \quad
  Jaejun Yoo\textsuperscript{1}\thanks{Corresponding author.} \\
  \textsuperscript{1}Ulsan National Institute of Science and Technology (UNIST) \\
  \textsuperscript{2}LOTTE INNOVATE \\
  \{rlagkdus705, softjin, jaejun.yoo\}@unist.ac.kr \\
}

\begin{document}

\maketitle

\begin{abstract}
Audio-video (AV) editing aims to modify audio and video content according to a target prompt. Unlike single-modality editing, AV editing requires models to infer a \emph{modality-selective edit scope} from the prompt alone: determining not only what should change, but also which modality should be preserved. Faithfully evaluating such models therefore requires both (i) benchmarks that span diverse edit types and modality categories, and (ii) evaluation that is itself modality-aware and sample-specific. However, existing AV editing benchmarks provide limited coverage of edit types and modality combinations, while current evaluation systems are often modality-blind and sample-agnostic, making it difficult to assess whether models faithfully preserve the unintended modality. To address these gaps, we introduce \ours, \textbf{A}udio-\textbf{V}ideo \textbf{E}diti\textbf{N}g \textbf{U}nderstanding and \textbf{E}valuation, comprising two contributions: (1) a benchmark of 1,291 source clips and 7,957 editing instructions across audio-targeted, video-targeted, and
AV-coupled edit types, curated and human-verified from VGGSound; and (2) a sample-specific, modality-aware evaluation framework that specifies, for each sample, both the intended change and the content that must remain intact. We evaluate representative AV editing models spanning three editing paradigms---joint, sequential, and separate---providing the first systematic analysis of modality-selectivity across paradigms. Our findings reveal a fundamental open challenge: when editing one modality, existing models frequently induce unintended changes in the other, regardless of paradigm. \ours~provides a benchmark and modality-aware evaluation framework to drive progress toward more controllable AV editing models. 
Our benchmark is publicly available at : \url{https://lait-cvlab.github.io/AVENUE-project-page/}.
\end{abstract}

\section{Introduction}

\begin{figure}[h]
    \centering
    \includegraphics[width=\linewidth]{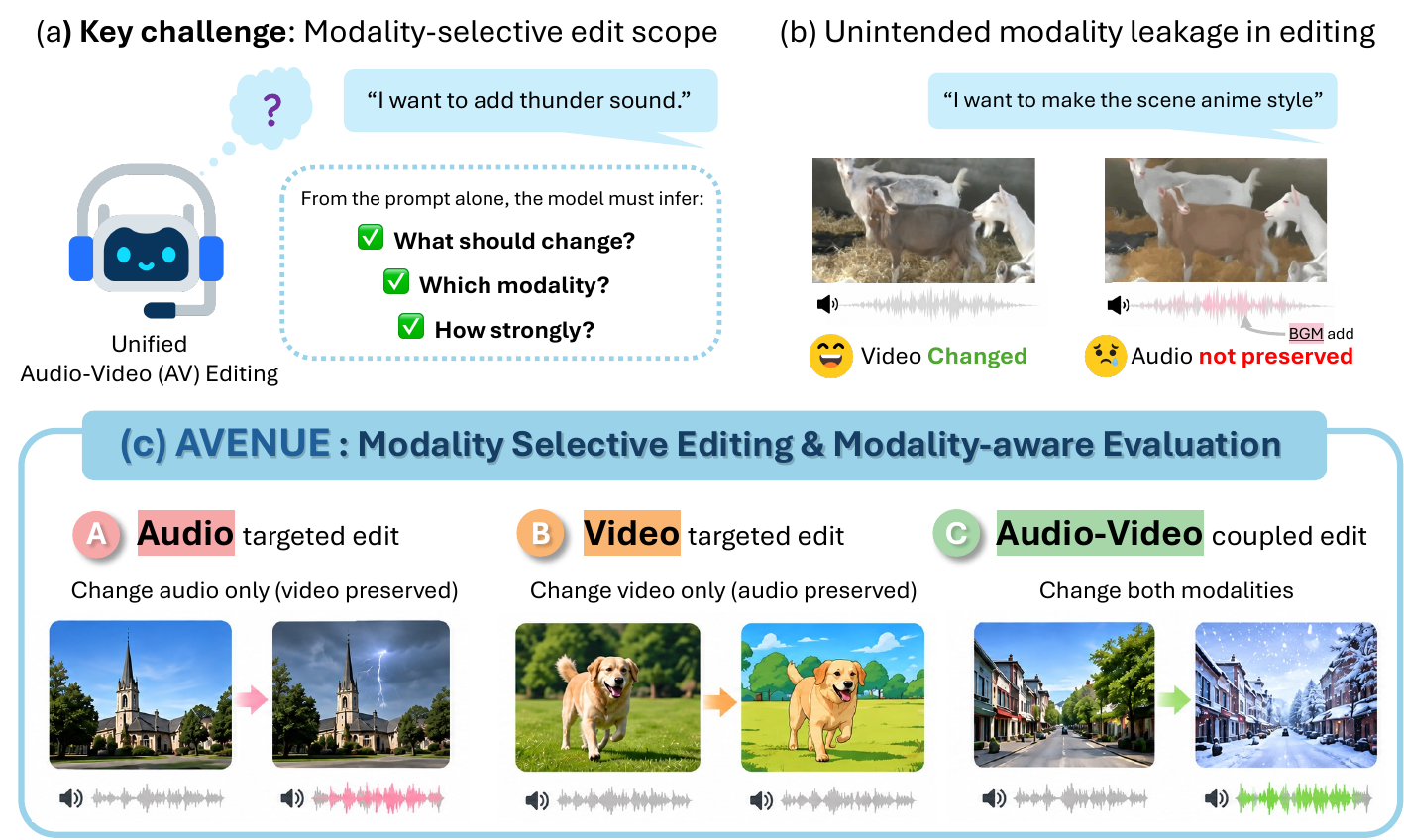}
    \caption{\textbf{Overview of AVENUE and the challenges of modality-selective editing.} (a) AV editing requires models to infer what to change, which modality to edit, and how strongly to modify it from the prompt alone. (b) Existing unified AV editing models may introduce unintended changes to the non-target modality, even when the intended edit succeeds. (c) AVENUE addresses these challenges through three editing categories—\textcolor{Audiocolor}{\textcircled{A} Audio-targeted},
    \textcolor{Videocolor}{\textcircled{B} Video-targeted}, and
    \textcolor{AVcolor}{\textcircled{C} AV-coupled edits}—each specifying which modalities should be modified or preserved. }
    \label{fig:teaser}
 \vspace{-2em}
\end{figure}

Recent advances in multimodal generative models have extended their applications beyond content creation to a broad range of content editing tasks. Among them, audio-video (AV) editing aims to modify both audio and video content according to a given target prompt. While recent work has proposed unified AV editing frameworks with diverse architectures, the task poses a challenge that single-modality editing does not: a model must infer \textbf{modality-selective edit-scope} from the prompt alone---determining what should change, which modality to edit, and how strongly each modality should be modified or preserved, as illustrated in Figure~\ref{fig:teaser}(a). This asymmetry is pervasive in practice. For example, adding thunder sound requires changing only the audio while preserving the video; applying a cartoon style requires changing only the video while preserving the audio; and transforming a scene into a snowy setting requires changing both the audio and video. However, existing AV editing models often fail to maintain this distinction, introducing unintended changes to the non-target modality even when the intended edit is successfully applied, as illustrated in Figure~\ref{fig:teaser}(b).

These challenges also demand a benchmark that explicitly evaluates the core capabilities of AV editing models. Yet existing benchmarks remain limited in both edit-type diversity and modality coverage. In terms of edit-type diversity, prior work~\cite{aved} is largely confined to object-level manipulation---covering only 1.3\% of the taxonomy we define (110 samples; see the \boxblue{blue} region in \Tref{tab:edit_taxonomy})---leaving attributes such as speed, action, and scene change largely unaddressed. In terms of modality coverage, existing benchmarks address only AV-coupled edits (see \Tref{tab:edit_taxonomy}), overlooking audio-targeted and video-targeted edits entirely. This narrow scope makes it impossible to evaluate whether a model can selectively edit one modality while preserving the other, motivating the three editing categories illustrated in Figure~\ref{fig:teaser}(c): audio-targeted, video-targeted, and AV-coupled.

Beyond data limitations, existing evaluation frameworks are fundamentally inadequate for AV editing. Current automated evaluation approaches---whether based on Vision-Language Models (VLM) judges, Multimodal Large Language Models (MLLM) scoring rubrics, or hybrid metric pipelines~\cite{li2025five, chen2025ivebench, zhao2025envisioning}---apply a fixed, sample-agnostic prompt to all test cases, relying on in-context examples for calibration. This design has two critical limitations in the AV setting. First, it is \textit{modality-blind:} it assesses whether the edit looks plausible overall, but does not explicitly measure whether the unintended modality was faithfully preserved---the very property that distinguishes a unified AV editor from two independent single-modal models. Second, it is \textit{sample-agnostic:} without grounded per-sample criteria specifying what must remain intact, the judge cannot reliably determine whether a subtle unintended change in the preserved modality constitutes a failure. As a result, these frameworks cannot surface the key behavioral differences between joint, sequential, and separate AV editing models.

To address these limitations, we introduce \textbf{\ours}: \textbf{A}udio-\textbf{V}ideo \textbf{E}diti\textbf{N}g \textbf{U}nderstanding and \textbf{E}valuation---a comprehensive benchmark 
with broad edit-type coverage and a modality-aware, sample-specific evaluation 
framework. \ours~is built on a curated subset of VGGSound~\cite{chen2020vggsound}, from which 1,291 unique source clips are retained after a three-stage filtering pipeline to ensure quality, yielding \textbf{7,957 editing instructions} across 4 audio-targeted, 3 video-targeted, and 5 AV-coupled edit types. Each sample is equipped with human-verified annotations specifying what changes and what must remain intact, grounding our proposed \textbf{Selective Controllability} metric---which explicitly measures whether the unintended modality is faithfully preserved---alongside Edit Accuracy, Modality Selectivity, Perceptual Quality, and AV Consistency.

To summarize, our main contributions are:
\begin{enumerate}
\item We introduce \textbf{\ours}, comprising 1,291 unique source clips and 7,957 editing instructions instances spanning 12 edit types across audio-targeted, video-targeted, and AV-coupled settings, substantially broadening the source scale, edit-type coverage, and modality coverage of existing AV editing benchmarks.
\item We provide per-sample human-verified annotations that specify the intended edit and elements that must remain intact, functioning as pseudo-GT to ground automated MLLM judges with sample-specific context beyond what generic rubrics can capture.
\item We introduce Selective Controllability— explicitly measuring whether the unintended modality remains intact after editing---as a dedicated metric absent from prior benchmarks. Combined with Edit Accuracy, Modality Selectivity, Perceptual Quality, and AV Consistency, we define a modality-type-specific evaluation framework tailored to the distinct challenges of audio-targeted, video-targeted, and AV-coupled edits.
\item We conduct a comprehensive study of existing AV editing models across three paradigms : joint, sequential and separate. We provide diagnostic insights into their robustness to unintended cross-modal effects and directions for future development.
\end{enumerate}

\begin{table}[t]
  \centering
  \caption{\textbf{(Left) Comparison with existing audio-video editing benchmarks.} We compare each benchmark in terms of the number of editing instances, supported edit types, edit modality, and data accessibility. A, V, and AV denote audio-targeted, video-targeted, and audio-video coupled editing, respectively.
  \textbf{(Right) \ours~ taxonomy.}
    \ours~defines a three-part edit taxonomy:
    \textcolor{Audiocolor}{\textcircled{A} Audio-targeted},
    \textcolor{Videocolor}{\textcircled{B} Video-targeted}, and
    \textcolor{AVcolor}{\textcircled{C} AV-coupled edits}. The outer sectors further categorize the 12 fine-grained editing types considered across the three taxonomy groups. The \boxblue{blue}-highlighted portion denotes \aved~\cite{aved}’s coverage, primarily sounding-object replacement.}
  \label{tab:edit_taxonomy}
  \begin{minipage}[c]{0.57\linewidth}
    \scriptsize
    \renewcommand{\arraystretch}{1.2}
    \setlength{\tabcolsep}{2pt}
    \definecolor{ourrow}{HTML}{3A8FD9}
    \resizebox{\linewidth}{!}{
    \begin{tabular}{lccc@{}c}
    \toprule
    \thead{Benchmark} & \thead{\# Editing \\ Instances} 
      & \thead{Edit Types} & \thead{Edit\\Modality} 
      & \thead{Data\\Access} \\
    \midrule
    OAVE~\cite{oave} & 1,100 
      & add, adjust
      & AV & \xmark \\
    \midrule
    Object-AVEdit~\cite{objectavedit} & - 
      & \makecell[c]{add,\\remove, replace}
      & AV & \xmark \\
    \midrule
    VGG-Edit~\cite{avedit} & 450 
      & \makecell[c]{add,\\remove, replace}
      & A & \xmark \\
    \midrule
    \aved~\cite{aved} & 110 
      & replace
      & AV & \cmark \\
    \midrule
    \rowcolor{ourrow!10}
    \textbf{\ours} & \textbf{7,957}
      & \makecell[c]{12 edit-types (add,\\remove, replace, change ...)}
      & \makecell[c]{Audio\\Video \\AV}
      & \cmark \\
    \bottomrule
    \end{tabular}
    }
  \end{minipage}
  \hfill 
  \begin{minipage}[c]{0.40\linewidth}
    \centering
    \includegraphics[width=\linewidth]{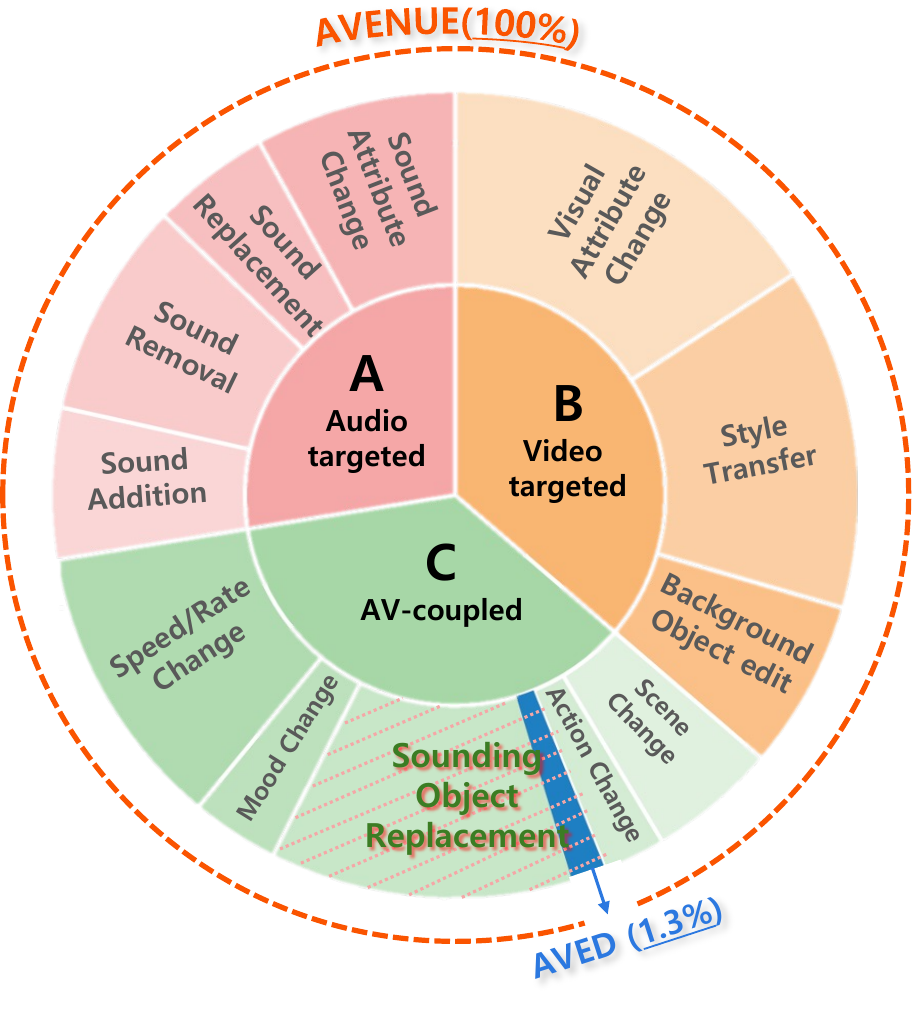}
 
  \end{minipage}
  \vspace{-2em}
\end{table}

\section{Related Works}
\paragraph{Single modal Generation and Editing.}
Recent advances in generative models have enabled text-guided content editing in both visual and audio domains~\cite{prompt-to-prompt, nulltextinversion, brooks2023instructpix2pix, zhao2025controlvideo, makeanaudio, audioldm, zeta, sdedit, rave}. In the visual domain, diffusion-based methods leverage pretrained latent diffusion models for prompt-guided image and video editing while preserving the structure of the source content~\cite{rombach2022high, sdedit, rave, prompt-to-prompt}. In the audio domain, latent diffusion models have similarly shown strong performance for text-conditioned audio generation and editing~\cite{audioldm, audioldm2, makeanaudio, zeta, smartdj, audit}. Although these methods are effective within their respective modalities, directly applying them independently to audio and video streams can break audio-visual coherence when the edited outputs no longer share consistent semantics. This limitation motivates unified AV editing models that explicitly coordinate edits across modalities~\cite{aved, coherent}.

\paragraph{Unified Audio-Video Editing and Benchmarks.}
To address the limitations of decoupled editing, recent works have explored unified AV editing frameworks that coordinate modifications across synchronized audio and visual streams~\cite{oave, aved, coherent, avedit}. These approaches differ in how they enforce cross-modal consistency: for example, some jointly optimize audio and video through shared or coupled latent guidance, while others follow a sequential paradigm that edits one modality conditioned on the other. Despite this progress, systematic evaluation of AV editing remains underdeveloped. Existing AV editing benchmarks are still limited in scale and edit diversity, often focusing on a small set of edit categories~\cite{oave, avedit, aved}, as summarized in \Tref{tab:edit_taxonomy}. Moreover, current evaluation protocols largely rely on embedding-based alignment metrics~\cite{radford2021learning, clap,girdhar2023imagebind}, which do not explicitly assess whether the intended modality is correctly edited while the unintended modality is preserved on a per-sample basis.

\paragraph{MLLM as-a-Judge.}
The use of foundation models as automated evaluators has gained traction in the assessment of generative models. Chen et al.~\cite{chen2024mllm} systematically investigate the capacity of MLLMs as judges across scoring, pair comparison, and batch ranking tasks, revealing that while MLLMs exhibit human-like discernment in pairwise comparisons, notable gaps remain in scoring and ranking, highlighting both the promise and current limitations of this paradigm. In the image generation domain, MMIG-Bench~\cite{hua2025mmig} employs a VQA-based metric leveraging vision-language models to assess fine-grained prompt-image alignment, demonstrating strong correlation with human judgments. Similarly, FiVE~\cite{li2025five} introduces FiVE-Acc, a VLM-based metric for evaluating object-level success in fine-grained video editing, further validating the utility of foundation models as scalable evaluators. Despite these advances, the application of MLLM-as-a-judge to audio-video editing evaluation remains largely unexplored, where the added complexity of cross-modal consistency demands more comprehensive automated assessment beyond unimodal metrics.

\section{\ours-Bench}
\label{dataset}

A core observation motivating our design is that real-world editing tasks are inherently \emph{selective}: a user may wish to increase the volume of an engine sound without touching the video, change the color of a vehicle without altering its sound, or replace a cat with a dog where both the video appearance and associated sound must change coherently. In other words, not all edits require modifying both modalities; unnecessary changes to an irrelevant modality indicate a lack of modality-aware understanding and control over the intended edit scope. 
This stands in contrast to prior benchmarks that treat audio-video editing as a monolithic task. To capture this diversity, we organize our benchmark along two axes: \emph{modality scope} (Audio-targeted, Video-targeted, or AV-Coupled) and \emph{semantic category} (e.g., Instruments, Animals, Human), and curate a dataset from VGGSound with taxonomy-aligned editing instructions for each combination. \Sref{sec:curation} describes the dataset construction pipeline, including the data collection and filtering procedures. \Sref{sec:taxonomy} presents an editing taxonomy organized across three modality-level groups. Finally, \Sref{sec:evaluation} introduces the evaluation framework used to assess model performance.

\subsection{Dataset Curation}
\label{sec:curation}

We construct \ours~through a rigorous multi-stage curation pipeline built on VGGSound~\cite{chen2020vggsound}, a large-scale audio-video dataset curated from YouTube videos with textual sound event labels. Our pipeline progressively filters raw clips and generates high-quality, taxonomy-aligned editing instructions. We source all clips exclusively from the \textbf{VGGSound test set} to avoid data leakage with models pretrained on VGGSound training data.

\paragraph{Category Filtering.} VGGSound labels span a wide range of sound sources. We group these labels into eight semantic categories: \textit{Instruments}, \textit{Animals}, \textit{Human}, \textit{Vehicles \& Engines}, \textit{Domestic \& Tools}, \textit{Sports \& Leisure}, \textit{Nature \& Environment}, and \textit{Weapons \& Explosions}. We exclude two groups from further processing: (1)~\textit{Speech}, because spoken language introduces linguistic content that lies outside the scope of our audio-video editing tasks, and (2)~\textit{Others}, a residual category containing labels that do not fit cleanly into any defined category and would introduce semantic ambiguity.

\paragraph{Audio-Video Quality Filtering.} We apply two-stage quality filtering: (1) {Audio-Video quality filtering} using ImageBind~\cite{girdhar2023imagebind} and CLAP~\cite{clap} scores, and (2) {Editing suitable filtering} comprising semantic consistency, person density, ROI consistency, and static video checks. Full details are provided in the App.~\ref{sec:dataset_curation}.

\subsection{Editing Taxonomy}
\label{sec:taxonomy}

Our taxonomy is organized into three modality-level groups shown in the right panel of \Tref{tab:edit_taxonomy}:

\textbf{A. Audio-targeted Edits.} Audio is targeted for editing; the video content must remain unchanged. Edit types include sound addition, removal, replacement, and intensity or pitch adjustment.

\textbf{B. Video-targeted Edits.} Video is targeted for editing; the audio must remain unchanged. Edit types include attribute change, style transfer, and background object manipulation.

\textbf{C. AV-coupled Edits.} Both modalities must change together in a semantically coherent manner. Edit types include sounding object replacement, scene/weather/mood changes, speed changes, and action changes.

A central design decision is that edit templates vary by semantic category. For example, \emph{Action Change} is meaningful for Human clips but inapplicable to Nature scenes; \emph{Weather Change} is meaningful for Nature and Animals but not for Instruments. To enforce this, we define a \emph{category-aware edit template} for each of the eight semantic categories shown in Figure~\ref{fig:pipeline}, specifying which edit types are valid. This prevents the generation of semantically inconsistent instructions and ensures that every benchmark sample is both realistic and evaluable.

\begin{figure}[t]
    \centering
    \includegraphics[width=\linewidth]{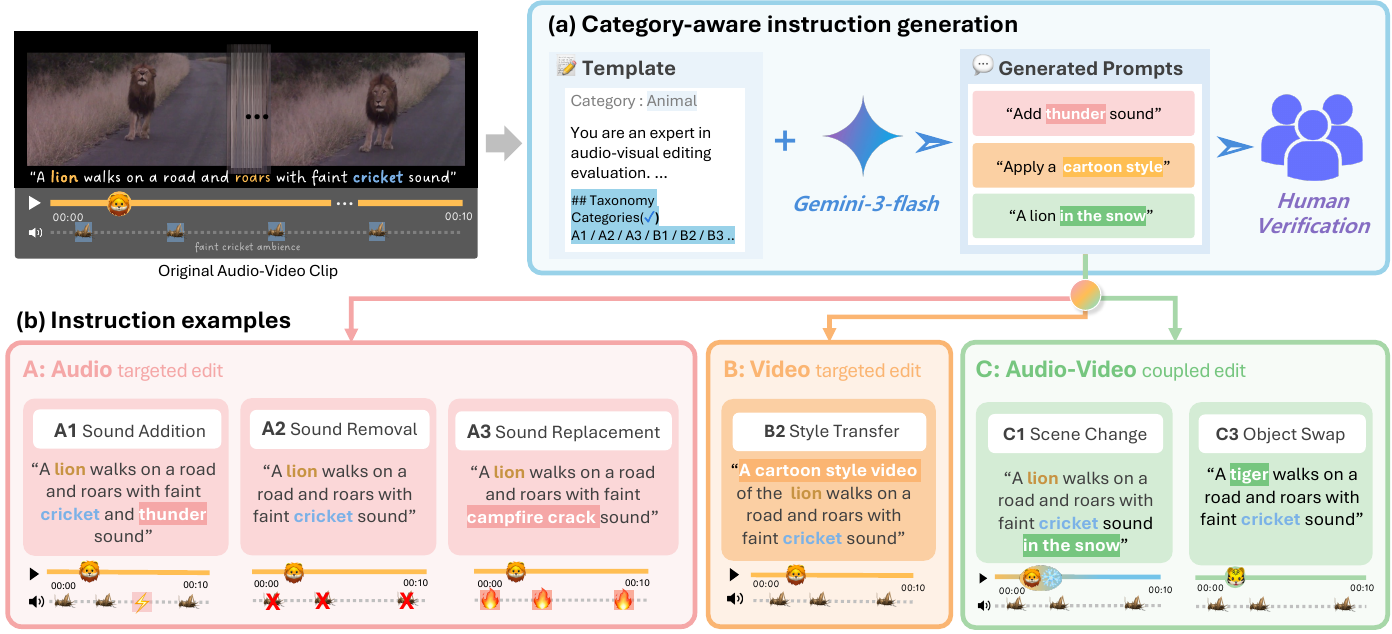}
    \caption{\textbf{Instruction process for AVENUE.} (a) Category-aware instruction generation using Gemini-3-flash~\cite{gemini36flash} with human verification, producing structured edit prompts for a source audio–video clip across 12 taxonomy categories. (b) Representative instruction examples across modality categories, with symbols below each target prompt illustrating the intended video and audio outcomes.
    }
    \label{fig:data_example}
\end{figure}

\paragraph{Editing Instruction Generation and Filtering.} 

For each retained clip, we generate editing instructions using \textbf{gemini-3-flash-preview}~\cite{team2023gemini} as illustrated it Figure~\ref{fig:data_example}(a). Rather than applying a single universal prompt, we supply each clip with a \emph{category-aware edit template} (see detail in \supple \ref{tab:taxonomy} and \ref{sup_sec:category_aware_editing_taxonomy}) that enumerates only the edit types valid for that category. This design ensures that generated instructions are semantically grounded and avoids producing edits that are inapplicable to the clip's content (e.g., generating an \emph{Action Change} instruction for a \emph{Nature} clip). The model is prompted to produce a concrete, specific instruction for each applicable edit type given the video and audio content of the clip. As shown in Figure~\ref{fig:data_example}(b), a single clip can yield diverse editing instructions spanning audio-targeted, video-targeted, and AV-coupled edits. Generated instructions are then filtered to retain only those that align with a valid taxonomy entry, discarding vague or malformed outputs.

\subsection{Selective Controllability Evaluation}

\begin{figure}[t]
    \centering
    \includegraphics[width=\linewidth]{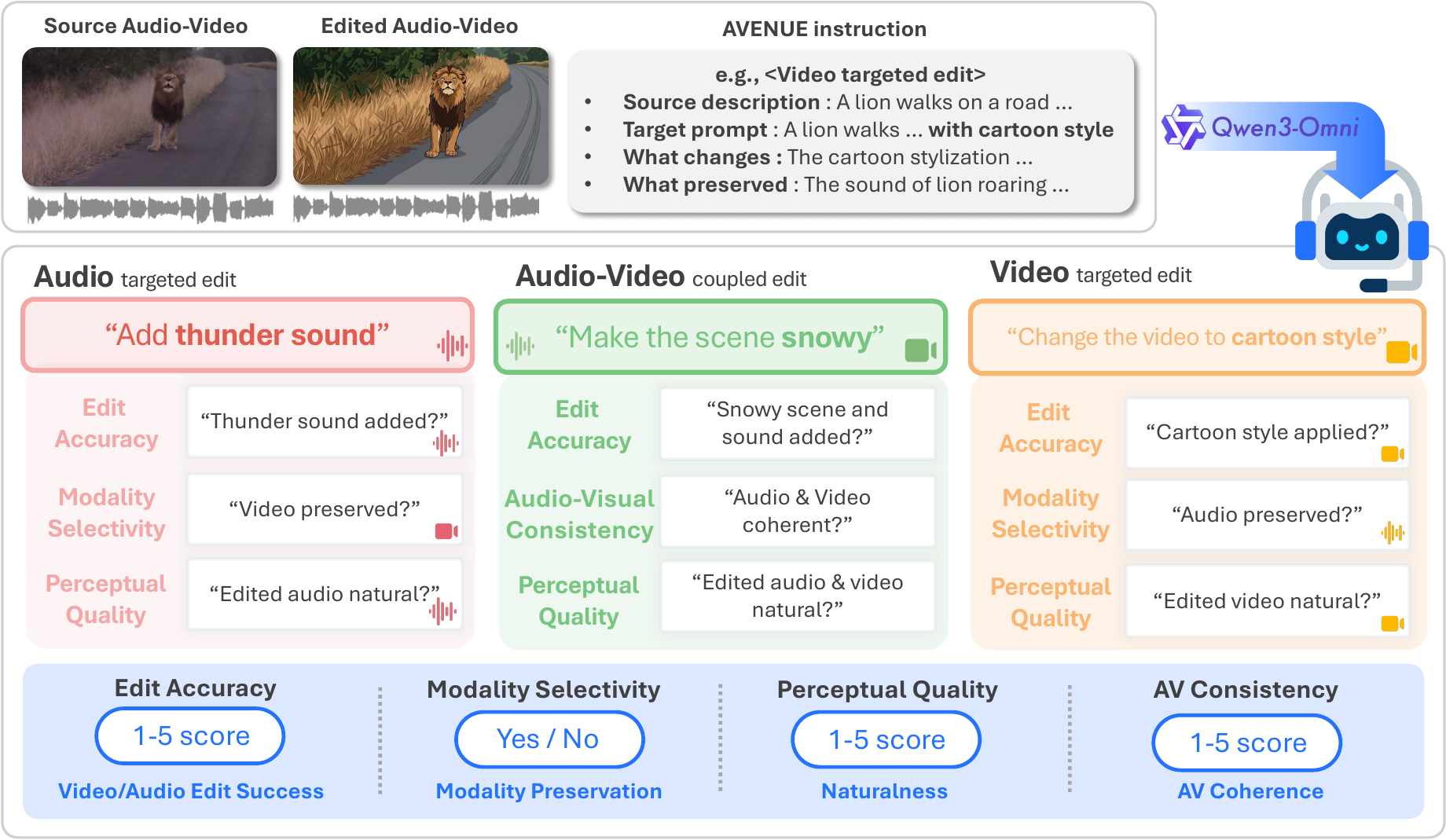}
    \caption{\textbf{Multimodal LLM-as-a-judge evaluation.} Qwen3-Omni~\cite{Qwen3-Omni} takes the source and edited audio–video pairs along with the AVENUE instruction, and performs sample-specific evaluation using the dimensions applicable to each modality category: Edit Accuracy, Modality Selectivity, Perceptual Quality, and AV Consistency. The input example illustrates a video-targeted edit: ``Change the video to cartoon style.''}
    \label{fig:evaluation_framework}
\end{figure}

\label{sec:evaluation}
  We evaluate audio-video editing quality across four complementary
  dimensions, as illustrated in ~\Fref{fig:evaluation_framework}. Rather than relying on a single score, our metrics  separately assess edit accuracy, modality selectivity, perceptual quality,
  and Audio-Video consistency, reflecting the multifaceted nature of audio-video editing. 
  To overcome the modality-blind and sample-agnostic limitations of             
  existing frameworks~\cite{li2025five, chen2025ivebench, zhao2025envisioning},
  we ground each evaluation in human-verified per-sample annotations:           
  \texttt{what\_changed}, specifying the intended modification, and             
  \texttt{what\_preserved}, specifying the content that must remain intact.
  These serve as pseudo-ground-truth criteria, enabling the evaluator to        
  assess each sample against its own explicitly defined edit scope.
  Following recent MLLM-as-judge approaches, we employ                           
  Qwen3-Omni-30B~\cite{Qwen3-Omni} as an evaluator which takes multimodal inputs.

\begin{itemize}[leftmargin=0.8em, itemsep=0pt, topsep=2pt]
  \item \textbf{Edit Accuracy (EA)} measures whether the target edit instruction has been faithfully applied within the targeted modality, conditioned on \texttt{what\_changed}. The evaluator is provided with both source and edited outputs and produces a score on a 1--5 Likert scale.

  \item \textbf{Modality Selectivity (MS)} evaluates whether the
  unintended modality remains intact after editing---e.g., video should be unchanged under an audio-targeted prompt. Both source and edited outputs are provided. Rather than a scalar score, this metric is assessed via five binary (Yes/No) questions covering general preservation criteria, plus one sample-specific question grounded in \texttt{what\_preserved}, directly addressing the modality-blind limitation of prior frameworks. EA and MS are grounded in disjoint fields (\texttt{what\_changed} vs.\ \texttt{what\_preserved}) and can dissociate: an edit may be correct (high EA) yet leak into the untargeted modality (low MS).
  
  \item \textbf{Perceptual Quality (PQ)} assesses the intrinsic quality and naturalness of the edited output, independent of the source. Only the edited output is provided to avoid conflating quality judgments with edit fidelity. Scored on a 1--5 Likert scale.
  
  \item \textbf{AV Consistency (AV-C)} measures the synchrony and semantic         
  coherence between the audio and visual streams of the edited
  output. Scored on a 1--5 Likert scale.                                    
\end{itemize}   

For all metrics, the input modality supplied to the evaluator is aligned with the edit category: audio-targeted for Category~A, video-targeted for Category~B, and audio-visual for Category~C, ensuring the evaluator attends to the relevant signal without modality-induced bias. Detailed evaluation prompts and input configurations are provided in the \supple \ref{sec:mllm}.

\section{Experiments} 
\label{experiments}
\subsection{Preliminaries}
\label{experiments-setup}
\subsubsection{Audio-Video Editing Paradigms} 

We analyze audio-video editing models under three paradigms that reflect 
distinct approaches to cross-modal integration.
Formally, let $v$, $a$, and $p$ denote the source video, source audio, and editing prompt, respectively, and let $f_V$, $f_A$, and $f_{AV}$ denote the video, audio, and joint editing functions.

\textbf{Separate paradigm} independently applies dedicated single-modality models for video and audio editing without cross-modal conditioning. We include RAVE+ZETA and RAVE+SDEdit, where RAVE~\cite{rave} handles video editing and ZETA~\cite{zeta} or SDEdit~\cite{sdedit} independently edits the audio track:
\begin{equation}
    \hat{v} = f_{V}(v,\, p), \qquad \hat{a} = f_{A}(a,\, p).
\end{equation}

\textbf{Sequential paradigm} adopts a video-to-audio (V2A) pipeline, where visual content is first edited and the resulting video serves as the condition for audio synthesis. We evaluate \cave~\cite{coherent} and MMAudio~\cite{mmaudio}. MMAudio is included as a generative reference that synthesizes audio solely from the edited visual stream, without conditioning on the source audio, whereas \cave~introduces source-audio conditioning to the MMAudio backbone. Their comparison therefore helps assess the effect of source-audio anchoring on editability and preservation.
\begin{equation}
    \hat{v} = f_{V}(v,\, p), \qquad \hat{a} = f_{A}(a,\, \hat{v},\, p).
\end{equation}

\textbf{Joint paradigm} processes audio and visual streams within a single unified framework, enabling direct cross-modal interaction during generation. We include \aved~\cite{aved}, which jointly edits both modalities conditioned on the target prompt:
\begin{equation}
    (\hat{v},\, \hat{a}) = f_{AV}(v,\, a,\, p).
\end{equation}

All models are evaluated across all three edit taxonomies (audio-targeted, video-targeted, and AV-coupled). To ensure a fair comparison, the sequential and separate paradigms share the same RAVE~\cite{rave}-edited video as input, isolating the effect of audio-video integration strategy from video editing capability. 

\subsubsection{Self-prompt preservation of AV Editing models}
\label{sec:self_aug}

A fundamental requirement for modality-selective editing is that a model alters only the target modality while leaving the other intact. To probe this, we construct a \textit{null-edit} setting where the source and target prompts are identical $(\mathrm{src} = \mathrm{trg})$
, under which a well-behaved model should produce output indistinguishable from the input. Any deviation indicates unintended cross-modal interference, independent of whether the intended edit was correctly applied. We construct 50 such pairs from Category~C and evaluate each model using the same modality-specific metrics.

Table~\ref{tab:self_prompt_mllm} reports preservation scores under the null-edit condition. \textbf{Separated} models, by design, enforce strict modality independence: the video and audio branches share no information, making cross-modal interference  structurally impossible. Their audio deviation under the null-edit condition (MS-Audio/CLAP: 48.3/0.721 for RAVE+SDEdit and 76.3/0.859 for RAVE+ZETA) itself, rather than contamination from the video stream. \textbf{Sequential} models exhibit similar audio instability (MS-Audio/CLAP: 58.7/0.736 for RAVE $\rightarrow$ CAVE and 56.3/0.731 for Source $\rightarrow$ CAVE), as their audio branch generates unconditionally from video without an anchor to the original audio signal. \textbf{Joint} achieves the best preservation scores across both modalities (MS-Video/V-CLIP~\cite{wang2024videoclip}: 76.3/0.995; MS-Audio/CLAP~\cite{clap}: 85.0/0.886 for AvED), indicating stronger null-edit robustness relative to the other paradigms. The consistent trends between MS and the embedding-based metrics further support the observed preservation behavior; additional embedding-based analyses are provided in \supple \ref{sec:self_prompt_embeddings}. However, this raises a critical question: \textit{does high AV Consistency reflect genuine editing quality, or merely a tendency to preserve the original input unchanged?} We analyze this further in Section~\ref{sec:cat_ab}.

\begin{table}[t!]
\centering
\caption{\textbf{Self-prompt preservation across AV editing model types.} We evaluate preservation under a null-edit condition where the target prompt matches the source prompt. We report our MLLM-based Modality Selectivity (MS) together with representative embedding-based preservation metrics~\cite{wang2024videoclip, clap}. Higher scores indicate better preservation. Shared video outputs (RAVE) are reported once across corresponding pipelines, and ``--'' denotes metrics that are not applicable. \textbf{Bold} and \underline{underlined} indicate the best and second-best results, respectively.}
\label{tab:self_prompt_mllm}
\vspace{2mm}
\begin{tabular}{c
    >{\centering\arraybackslash}p{1.6cm}
    >{\centering\arraybackslash}p{1.6cm}|cc|cc}
\toprule
\multirow{2}{*}{\textbf{Model Type}} & \multicolumn{2}{c}{\textbf{Model}}
  & \multicolumn{2}{c}{\textbf{Video Preservation}$\uparrow$}
  & \multicolumn{2}{c}{\textbf{Audio Preservation}$\uparrow$} \\
\cmidrule(lr){2-3} \cmidrule(lr){4-5} \cmidrule(lr){6-7}
& \textbf{Video} & \textbf{Audio}
  & \textbf{MS (Video)} & \textbf{V-CLIP}
  & \textbf{MS (Audio)} & \textbf{CLAP} \\
\midrule
\multirow{2}{*}{\textit{Separated}} & \multicolumn{2}{c|}{RAVE $+$ SDEdit}
  & \multirow{3}{*}{\underline{70.7}} & \multirow{3}{*}{\underline{0.956}} & 48.3 & 0.721 \\
& \multicolumn{2}{c|}{RAVE $+$ ZETA}
  &  &  & \underline{76.3} & \underline{0.859} \\
\cmidrule{1-3} \cmidrule{6-7}
\multirow{2}{*}{\textit{Sequential}} & \multicolumn{2}{c|}{RAVE $\rightarrow$ CAVE}
  &  &  & 58.7 & 0.736 \\
& \multicolumn{2}{c|}{Source $v \rightarrow$ CAVE}
  & \cellcolor{gray!40}-- & \cellcolor{gray!40}-- & 56.3 & 0.731 \\
\midrule
\textit{Joint} & \multicolumn{2}{c|}{AvED}
  & \textbf{76.3} & \textbf{0.995} & \textbf{85.0} & \textbf{0.886} \\
\bottomrule
\end{tabular}
\end{table}

\subsection{Comprehensive Analysis} 

\subsubsection{AV-coupled Edit Results}
\label{sec:cat_c}

\begin{figure}[t]
    \centering
    \includegraphics[width=\linewidth]{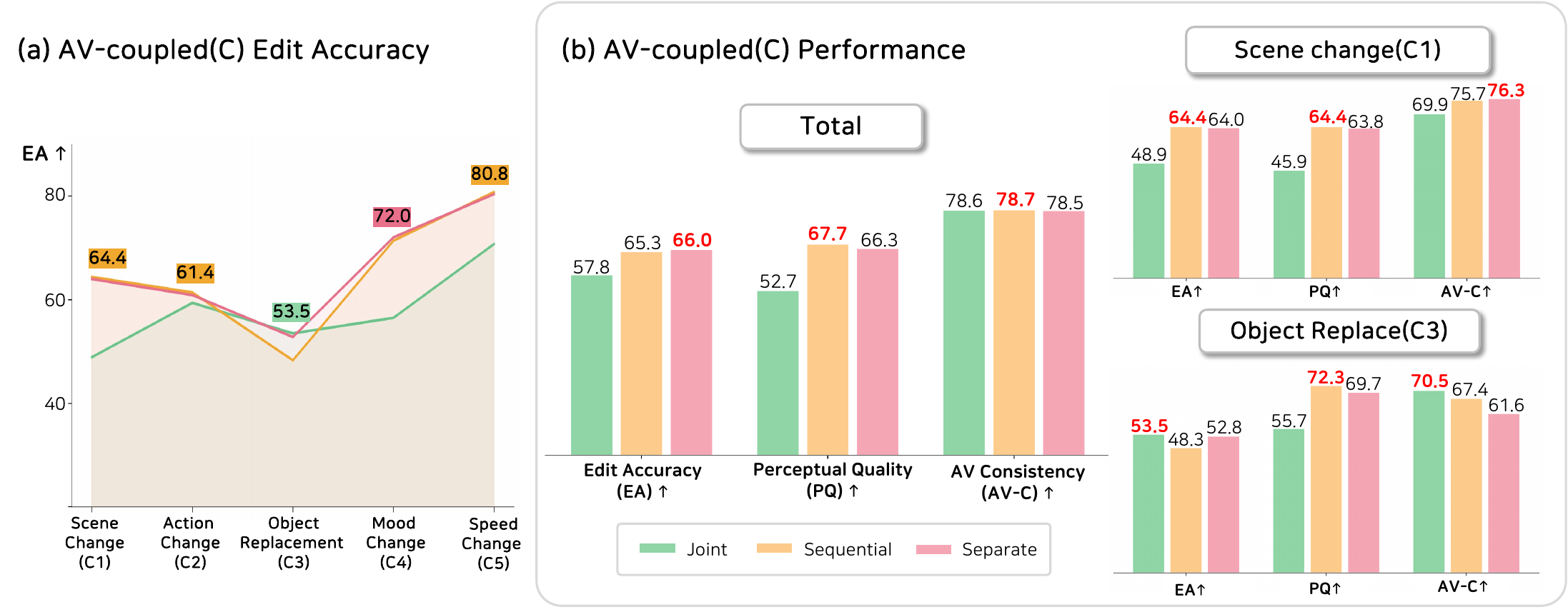}
    \caption{
    \textbf{Evaluation of AV-coupled editing under \cat C.} All scores are reported out of 100, where higher is better. (a) Edit Accuracy (EA) across five sub-categories of Category C. (b) Comparison of joint, sequential, and separate editing paradigms across EA, Perceptual Quality (PQ), and AV Consistency (AV-C). We report results on the full Category C benchmark as well as two representative cases: Scene Change (C1) and Sounding Object Replacement (C3).}
    \label{fig:c_perform}
\end{figure}

We first examine how editing paradigms perform across different edit sub-types within \cat~C (AV-coupled editing). As shown in Figure~\ref{fig:c_perform}(a), the Edit Accuracy of each paradigm varies considerably depending on the edit sub-type. Notably, the ranking among paradigms is not consistent: Sequential and Separate methods outperform Joint by a large margin on C1 and C4, yet Joint reverses this gap on C3. For C5(speed change), where the editing scope is narrow and the transformation is relatively straightforward, all three paradigms converge to similar performance, suggesting that paradigm differences are most pronounced for challenging edits. \\

Figure~\ref{fig:c_perform}(b) summarizes the aggregate performance on Category~C across three metrics.                                                          Interestingly, AV Consistency scores are nearly identical across all paradigms ($\sim$78.5), indicating that cross-modal coherence is comparably achievable regardless of architectural design. The key differentiators are Edit Accuracy, where Joint trails Sequential and Separate by 7--8 points, and Perceptual Quality, where the gap widens to $\sim$15 points. Sequential methods achieve the most balanced profile overall, while Separate methods show a slight edge in Edit Accuracy. \\                           

Zooming into C1 (Figure~\ref{fig:c_perform}(c)), Joint achieves the lowest scores across all three metrics (Edit Accuracy: 48.9, Perceptual Quality: 45.9, AV Consistency: 69.9), falling behind Sequential and Separate by 15--18 points. This sub-type involves fine-grained attribute modifications where independent, specialized modules can precisely target changes in each modality without interference, an advantage that the joint paradigm cannot leverage. \\

In contrast, C3 represents the most challenging sub-type, where all paradigms achieve their lowest scores (Figure~\ref{fig:c_perform}(d)). Here, Joint ranks first in both Edit Accuracy (53.5) and AV Consistency (70.5), suggesting that \textit{holistic cross-modal understanding} becomes critical when the edit involves replacing the sounding object itself---a task that inherently requires simultaneous reasoning over both modalities. However, Perceptual Quality remains in favor of Sequential (72.3 vs.\ 55.7), revealing that the generation quality bottleneck of joint models persists even in their most favorable setting. \\
Taken together, these results reveal that current AV editing methods exhibit complementary strengths across different edit types---from fine-grained attribute changes to holistic object replacements. Rather than favoring a narrow subset of edits, a comprehensive benchmark must span this full spectrum to faithfully assess model capabilities. We argue that this diversity is not a limitation but a \textit{desideratum}: real-world AV editing demands models that gracefully handle the entire range of edit granularities, and our benchmark is designed to expose where current paradigms fall short of this goal, guiding future research toward more versatile and robust AV editing systems.  

\subsubsection{Results on Audio-targeted and Video-targeted Editing}
\label{sec:cat_ab}

\begin{table*}[t]                                                                                                          
\small
\centering                                                                                                                                                                                               
\caption{\textbf{Audio modality evaluation overview.} All scores are scaled to 100. Edit Accuracy and Perceptual Quality are measured on Cat.A (audio-targeted) samples, and Modality Selectivity on Cat.B (video-targeted) samples to assess unintended audio change. Higher is better for all metrics.}
\label{tab:audio_overview}                                                                                                                                                                               
\begin{tabular}{l c@{\,}c@{\,}c ccc}                                                                                                                                                                     
\toprule                                                                                                                                               
\multirow{2}{*}{\textbf{Model Type}} & \multicolumn{3}{c}{\textbf{Model}} & \textbf{Edit Accuracy} & \textbf{Percept Quality} & \textbf{Modal Selectivity~(Audio)} \\
\cmidrule(lr){2-4} \cmidrule(lr){5-5} \cmidrule(lr){6-6} \cmidrule(lr){7-7} & \textbf{Video} & & \textbf{Audio} & (Cat A)$\uparrow$ & (Cat A)$\uparrow$ & (Cat B)$\uparrow$ \\ 
\midrule
\multirow{2}{*}{\textit{Separate}}
& RAVE & $+$ & ZETA & 79.7 & 88.2 & 93.0 \\                                                                                 & RAVE & $+$ & SDEdit & 78.7 & 82.4 & 51.4 \\
\midrule                                                                                                                    \multirow{4}{*}{\textit{Sequential}}                                                                                               & RAVE & $\rightarrow$ & CAVE & 79.6 & 89.3 & 58.0 \\ 
& Source $v$& $\rightarrow$ & CAVE & \textbf{79.8} & \textbf{89.7} & -- \\[2pt]
\cdashline{2-7}
\noalign{\vskip 2pt}
& RAVE & $\rightarrow$ & MMAudio & 75.8 & 89.5 & 39.7 \\
& Source $v$& $\rightarrow$ & MMAudio & 77.6 & 88.8 & -- \\
\midrule                                                                                                                                                                                                 
\textit{Joint} & \multicolumn{3}{c}{AvED} & 64.1 & 87.7 & \textbf{97.5} \\
\bottomrule                                                   \end{tabular}
\end{table*}

In Section~\ref{sec:self_aug}, we observed that AVED achieves notably high scores under the self-prompt preservation setting (source $=$ target), raising the question of \textit{whether this reflects genuinely robust editing or a tendency to leave the input largely unchanged.} The results in this section provide a clear answer. Table~\ref{tab:audio_overview} summarizes the audio modality evaluation across all paradigms. To ensure a fair comparison of editing paradigms, we fix RAVE as the shared video backbone across all experiments, as AVED's video component also follows the RAVE architecture based on Stable Diffusion 2.1. AVED achieves the highest modality selectivity (97.5), yet records the lowest edit accuracy (64.1) among all models. This confirms that AVED's high self-prompt preservation scores reflect a tendency to preserve rather than an ability to edit.

In contrast, separated and sequential paradigms achieve substantially higher edit accuracy (76–80), though their modality selectivity varies dramatically depending on the audio model. Under the same RAVE video backbone, ZETA attains 93.0 in audio preservation while SDEdit scores only 51.4—a gap driven entirely by the choice of audio model. The source $v$ ablation in sequential pipelines further supports this: bypassing video editing (source $v$ $\rightarrow$ CAVE vs.\ RAVE $\rightarrow$ CAVE) yields nearly identical scores across all metrics, indicating that upstream video editing introduces minimal accumulated error. These results suggest that for audio modality, the dominant factor in both editing quality and modality preservation is the audio model itself, rather than the video editing stage.

To understand what drives these differences, we examine the underlying mechanism of each audio model. \textbf{First}, ZETA, an inversion-based model, achieves strong performance in both edit accuracy (79.7) and modality selectivity (93.0). By inverting the input audio into its latent representation, ZETA preserves the source audio structure and modifies only the targeted elements. SDEdit, in contrast, perturbs the input with Gaussian noise and denoises toward a new text prompt. While this achieves comparable edit accuracy (78.7), preservation drops substantially (51.4), as the noising step inevitably loses part of the source information. \textbf{Second}, MMAudio, as a generation model, shows a distinctive pattern: perceptual quality remains high (89.5) since it synthesizes coherent audio from scratch, yet modality selectivity drops to the lowest (39.7) as it has no mechanism to retain the source audio. Notably, the automated judge correctly captures these structural characteristics of each model type, suggesting that the MLLM-based evaluation reflects meaningful differences in model behavior beyond surface-level quality.

These results highlight that the audio model's underlying mechanism—whether inversion-based, noise-based, or generative- is the key factor determining both editing and preservation performance. No single paradigm dominates all metrics: joint editing excels at preservation but under-edits, while noise-based and generative approaches achieve stronger edits but fail to preserve unedited content. Among the evaluated models, the inversion-based approach (ZETA) is the only one that achieves competitive performance across both axes. As audio-visual editing matures, inversion-based mechanisms may offer a viable path toward modality-aware editing that selectively modifies one modality without degrading the other.

\subsection{Human validation of the automated judge.}

To validate our MLLM-based evaluation against human judgments, we conducted a human--judge alignment study involving 13 human annotators, with detailed results reported in Table~\ref{tab:judge_agreement}. Our primary judge, Qwen3-Omni~\cite{Qwen3-Omni}, achieved 67.0\% pairwise agreement with humans, close to the 71.1\% human--human reference, and showed consistent trends under leave-one-out agreement (72.5\% vs. 79.2\%) and ranking-level Kendall's $\tau_b$ (0.353 vs. 0.546). Evaluating four additional omnimodal judges yielded similar trends, with Gemini-3.1-Pro showing comparable alignment and achieving higher agreement on some metrics. These results support Qwen3-Omni as a reliable primary evaluator and indicate that the observed human alignment is not specific to a single judge model. Full protocols and analyses are provided in \supple \ref{sec:human_alignment_judge}, with additional comparisons to embedding-based metrics in \supple \ref{supsec:embedding_based_eval}.

\begin{table}[h]
\centering
\footnotesize
\caption{\textbf{Judge agreement with human judgments.} We report individual-level pairwise agreement, leave-one-out agreement, and Kendall's $\tau_b$ for five automated judges. \textit{Human reference} denotes the corresponding human--human agreement, averaged across the five judge-specific human references. \textbf{Bold} and \underline{underline} denote the best and second-best results among automated judges, respectively.}
\label{tab:judge_agreement}
\vspace{2mm}
\begin{tabular}{lccc}
\toprule
\textbf{MLLM-Judge} & \textbf{\shortstack{Pairwise agreement}}$\uparrow$
  & \textbf{Leave-one-out}$\uparrow$ & \textbf{Kendall $\tau$-b}$\uparrow$ \\
\midrule
\textit{Human reference} & \textit{71.1\%} & \textit{79.2\%} & \textit{0.546} \\
\midrule
Qwen3-Omni~\cite{Qwen3-Omni}       & \textbf{67.0\%}    & \underline{72.5\%} & 0.353 \\
Gemini-3.1-pro~\cite{gemini31pro}   & \underline{66.2\%} & \textbf{73.6\%}    & \textbf{0.440} \\
Gemini-3.6-Flash~\cite{gemini36flash} & 64.0\%             & 67.7\%             & \underline{0.359} \\
Qwen2.5-Omni~\cite{Qwen2.5-Omni}     & 60.4\%             & 64.8\%             & 0.092 \\
Audio-Visual-Flamingo~\cite{avflamingo}       & 56.8\%             & 58.7\%             & 0.117 \\
\bottomrule
\end{tabular}
\end{table}

\section{Conclusion}

We presented \textbf{\ours}, a benchmark and evaluation framework for audio-video editing, enabling the first systematic assessment of modality-selective editing across diverse edit types and model paradigms. Our evaluation reveals that no existing paradigm fully resolves this challenge: each exhibits a distinct trade-off between edit fidelity and modality preservation, and performance varies considerably across edit types — suggesting that no single architectural design generalizes across the full spectrum of AV editing. Among the evaluated models, inversion-based mechanisms show the most promise for achieving competitive performance on both axes. Taken together, these findings indicate that modality-selective controllability remains a fundamental open challenge, and that future AV editing models should be designed with explicit awareness of modality scope — knowing not only what to change, but what to leave intact.

\begin{ack}
This work was supported by the National Research Foundation of Korea (NRF) grant funded by the Korea government (MSIT) (RS-2022-NR071940), the Institute of Information \& communications Technology Planning \& Evaluation (IITP) grant funded by the Korea government (MSIT) (No.RS-2020-II201336, Artificial Intelligence Graduate School Program (UNIST)), (No.RS-2022-II220264, Comprehensive Video Understanding and Generation with Knowledge-based Deep Logic Neural Network), (RS-2022-00143911, AI Excellence Global Innovative Leader Education Program), (RS-2025-25442149).
\end{ack}

{
\small
\bibliographystyle{unsrtnat}
\bibliography{main}
}



\newpage
\appendix
\onecolumn
\section*{\centering AVENUE: Audio-Video EditiNg Understanding and Evaluation \\
Appendix}
\setcounter{section}{0} 
\renewcommand{\thesection}{\Alph{section}} 

\renewcommand{\thefigure}{\textbf{A}\arabic{figure}}
\setcounter{figure}{0}
\renewcommand{\thetable}{\textbf{A}\arabic{table}}
\setcounter{table}{0}

\section{\ours~Dataset Details}
\subsection{Dataset Access}
\label{supsec:dataset_access}
We provide \ours~ dataset annotation file and evaluation code. 
These are available for download from the following links: (\href{https://lait-cvlab.github.io/AVENUE-project-page/}{\textcolor{magenta}{LINK}}). 
Detailed dataset statistics are discussed in \Sref{supsec:dataset_statistics}.

\subsection{Dataset Statistics for~\ours}
\label{supsec:dataset_statistics}

\paragraph{Dataset Scale Details.} \ours~contains 1,291 unique source clips and 7,957 editing instruction evaluation instances. These two statistics characterize different aspects of the benchmark: the former reflects the scale of unique source content, while the latter represents distinct instruction-conditioned editing cases. Each source clip is associated with an average of 6.16 editing instructions (median 6, min 3, max 9, $\sigma=1.58$).

Multiple instructions are assigned to a source clip when the same content supports semantically distinct editing operations, such as adding or removing a sound, modifying a visual attribute, or jointly editing the audio and video. We therefore report benchmark results at the video--instruction level, since successful editing for one instruction does not necessarily imply successful editing for another. Evaluating multiple instructions on the same source also enables comparisons of different editing behaviors while controlling for source-content difficulty. In terms of unique source content, \ours~contains substantially more source clips than existing AV editing benchmarks (OAVE: 44, AvED: 110, and VGG-Edit: 450).

\begin{figure}[t!]
    \centering
    \includegraphics[width=0.5\linewidth]{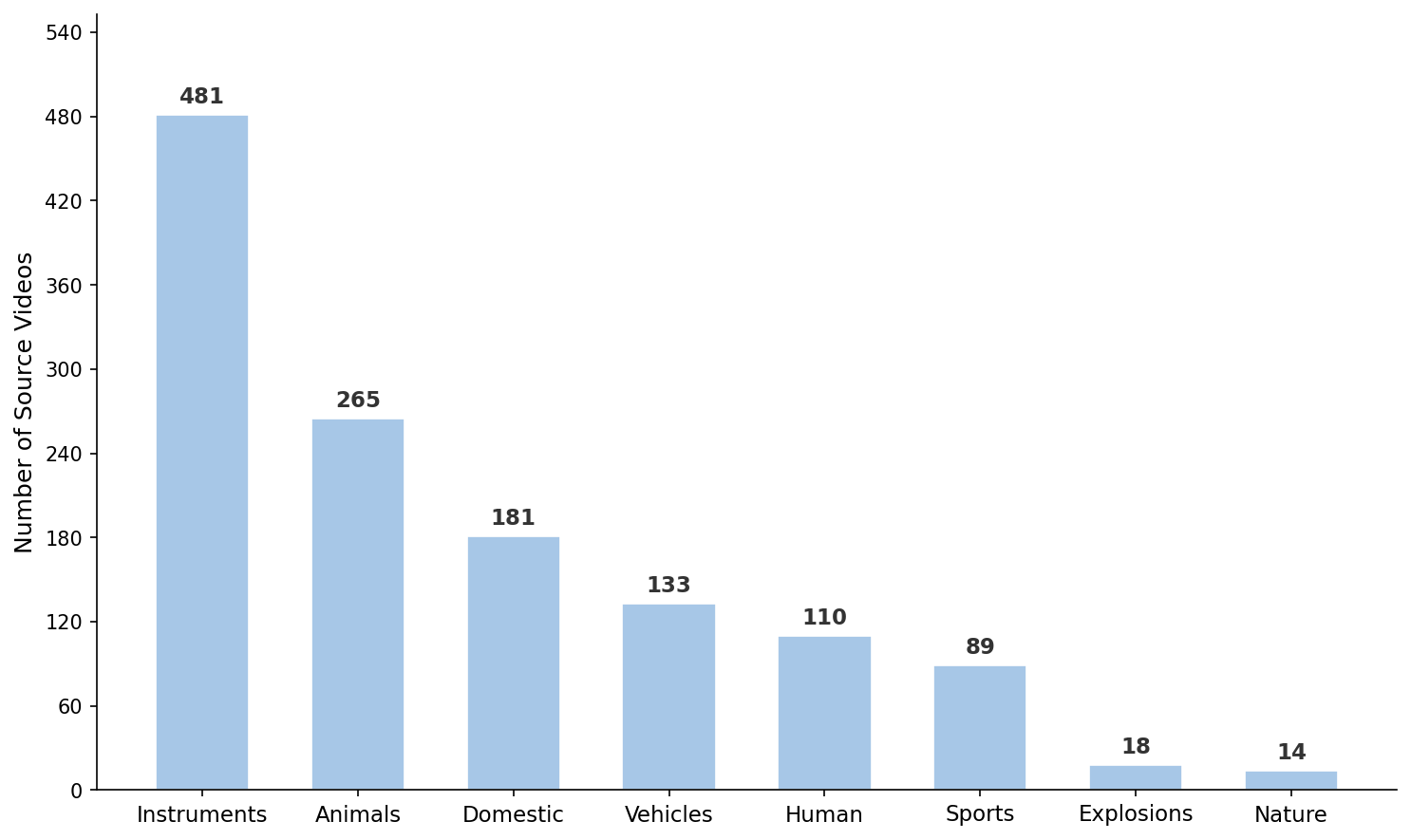}
    \caption{Statistics for semantic categories}
    \label{supfig:semantic_cat}
\end{figure}

\paragraph{Statistics for Semantic Category.} We define eight semantic categories to enable category-aware instruction generation, as described in main \Sref{sec:curation}. \Fref{supfig:semantic_cat} shows the distribution of source videos across these categories. Instruments dominates with 481 clips, followed by Animals (265) and Domestic (181), while Nature (14) and Explosions (18) are relatively underrepresented. This imbalance naturally reflects the underlying distribution of VGGSound~\cite{chen2020vggsound}, from which all clips are sourced.

\begin{figure}[h]
    \centering
    \begin{subfigure}[t]{0.32\linewidth}
        \centering
        \includegraphics[width=\linewidth]{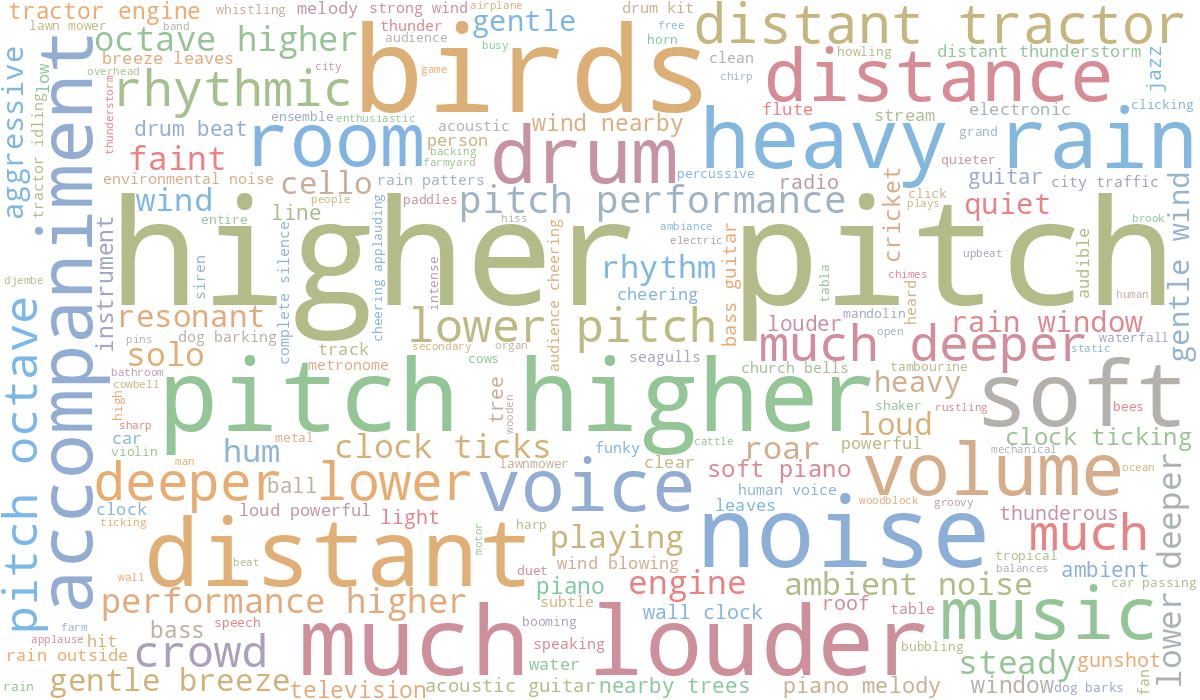}
        \caption{Audio-targeted edits (A1–A4)}
        \label{supfig:wc_audio-only}
    \end{subfigure}
    \hfill
    \begin{subfigure}[t]{0.32\linewidth}
        \centering
        \includegraphics[width=\linewidth]{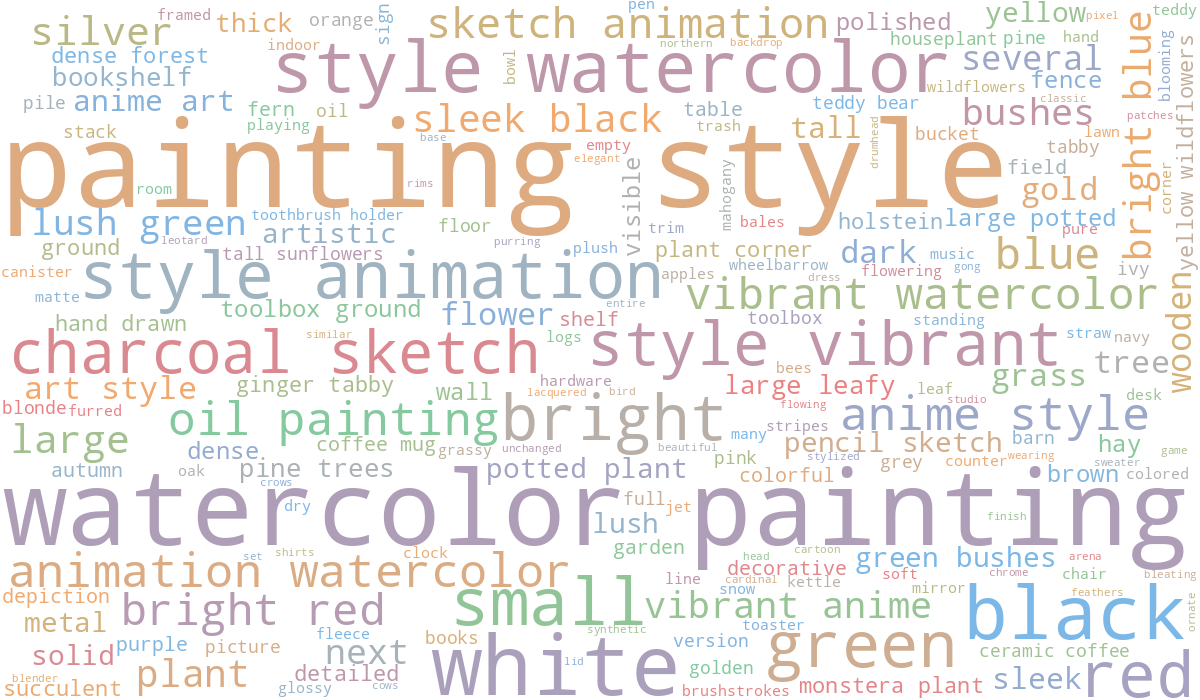}
        \caption{Visual-targeted edits (B1–B3)}
        \label{supfig:wc_video-only}
    \end{subfigure}
    \hfill
    \begin{subfigure}[t]{0.32\linewidth}
        \centering
        \includegraphics[width=\linewidth]{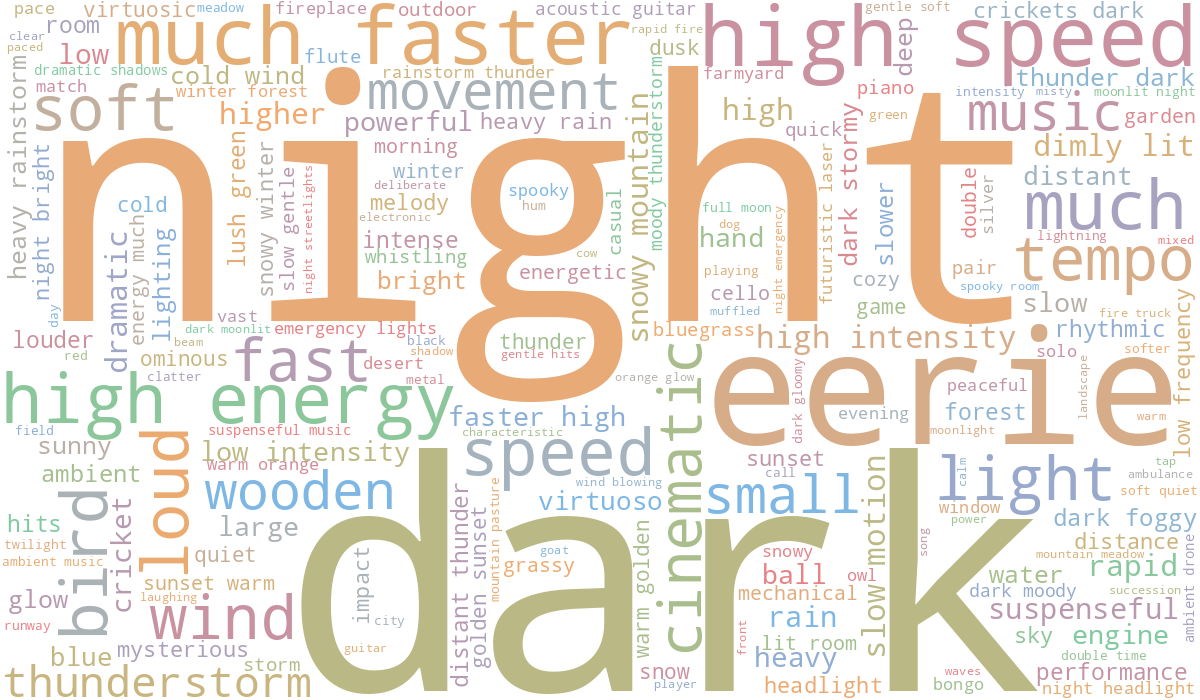}
        \caption{AV targeted edits (C1–C5)}
        \label{supfig:wc_cross-modal}
    \end{subfigure}

    \caption{Word clouds of content words added in editing prompts, grouped by editing modality: audio-targeted (A), visual-targeted (B), and AV-coupled (C). Words are extracted by aligning source descriptions and target prompts via sequence matching, then filtered to retain nouns and adjectives targeted, excluding domain-generic terms.}
    \label{supfig:word_cloud}
\end{figure}

\paragraph{Word Distribution for \ours.} In this section, we verify the diversity of edit instructions in our dataset by visualizing the semantic distribution of editing vocabulary through word clouds constructed separately for each edit category. As shown in \Fref{supfig:word_cloud}, the edit-specific terms are distinctly distributed across audio-targeted, video-targeted, and AV-targeted categories, confirming that our dataset covers a broad and varied range of editing semantics rather than converging on a narrow set of expressions.

\paragraph{Prompt Diversity Analysis.} 
\label{prompt_diversity_analysis}
To quantitatively assess the diversity of editing prompts in \ours, we compare \ours~ with the existing AV editing dataset AvED from two complementary perspectives: lexical diversity and semantic diversity. For lexical and surface-form diversity, we use Type-Token Ratio (TTR)~\cite{ttr}, distinct-(n)~\cite{distinct_n}, and self-BLEU~\cite{self_belu}, which measure vocabulary usage, (n)-gram diversity, and surface-form redundancy, respectively. For semantic diversity, we compute the mean pairwise cosine similarity between sentence embeddings extracted using all-MiniLM-L6-v2\footnote{\url{https://huggingface.co/sentence-transformers/all-MiniLM-L6-v2}}. Higher TTR and distinct-(n) indicate greater diversity, whereas lower self-BLEU and lower pairwise cosine similarity indicate less redundancy and greater diversity. Since these diversity statistics can be affected by corpus size, we perform a size-controlled comparison. Specifically, we randomly sample 110 prompts from \ours~ to match the number of prompts in AvED and repeat this procedure 200 times, reporting the mean value across the resulting subsets.

 As shown in Table~\ref{suptab:prompt_diversity}, \ours~ consistently exhibits greater prompt diversity than \aved. \ours~ achieves a TTR of 0.305, compared with 0.078 for \aved, and substantially higher distinct-1 (0.305 vs. 0.078) and distinct-2 (0.640 vs. 0.127). The unique-prompt ratio is also approximately 100\% for \ours, compared with 20.9\% for \aved. Moreover, \ours~ yields a considerably lower self-BLEU score (0.240 vs. 0.867), indicating substantially less repetition in surface expressions. This tendency also holds at the semantic level. \ours~ obtains a lower mean pairwise cosine similarity (0.160) than \aved~ (0.235), indicating that its prompts are not only lexically diverse but also span a broader range of semantic editing requests. Overall, these results show that \ours~ provides substantially greater diversity in both the linguistic formulation and semantic content of its edit instructions.

\begin{table}[t]
\centering
\caption{Size-controlled comparison of editing-prompt diversity between \ours~ and AvED.
\ours~ is randomly subsampled to $N=110$ to match the size of AvED, and results are
averaged over 200 random subsets. $\uparrow$ / $\downarrow$ indicate whether a higher
or lower value corresponds to greater diversity.}
\label{tab:prompt_diversity}
\begin{tabular}{lcc}
\toprule
Metric & \ours ($N{=}110$) & AvED ($N{=}110$) \\
\midrule
TTR ($\uparrow$)                    & \textbf{0.305}       & 0.078   \\  
distinct-2 ($\uparrow$)             & \textbf{0.640}       & 0.127   \\
Unique-prompt ratio ($\uparrow$)    & \textbf{$\sim$100\%} & 20.9\%  \\
self-BLEU ($\downarrow$)            & \textbf{0.240}       & 0.867   \\
Mean pairwise cosine ($\downarrow$) & \textbf{0.160}       & 0.235   \\
\bottomrule
\end{tabular}
\label{suptab:prompt_diversity}
\end{table}

\section{Dataset Curation}
\label{sec:dataset_curation}
\subsection{Dataset Filtering}
\begin{figure}[t!]
    \centering
    \includegraphics[width=\linewidth]{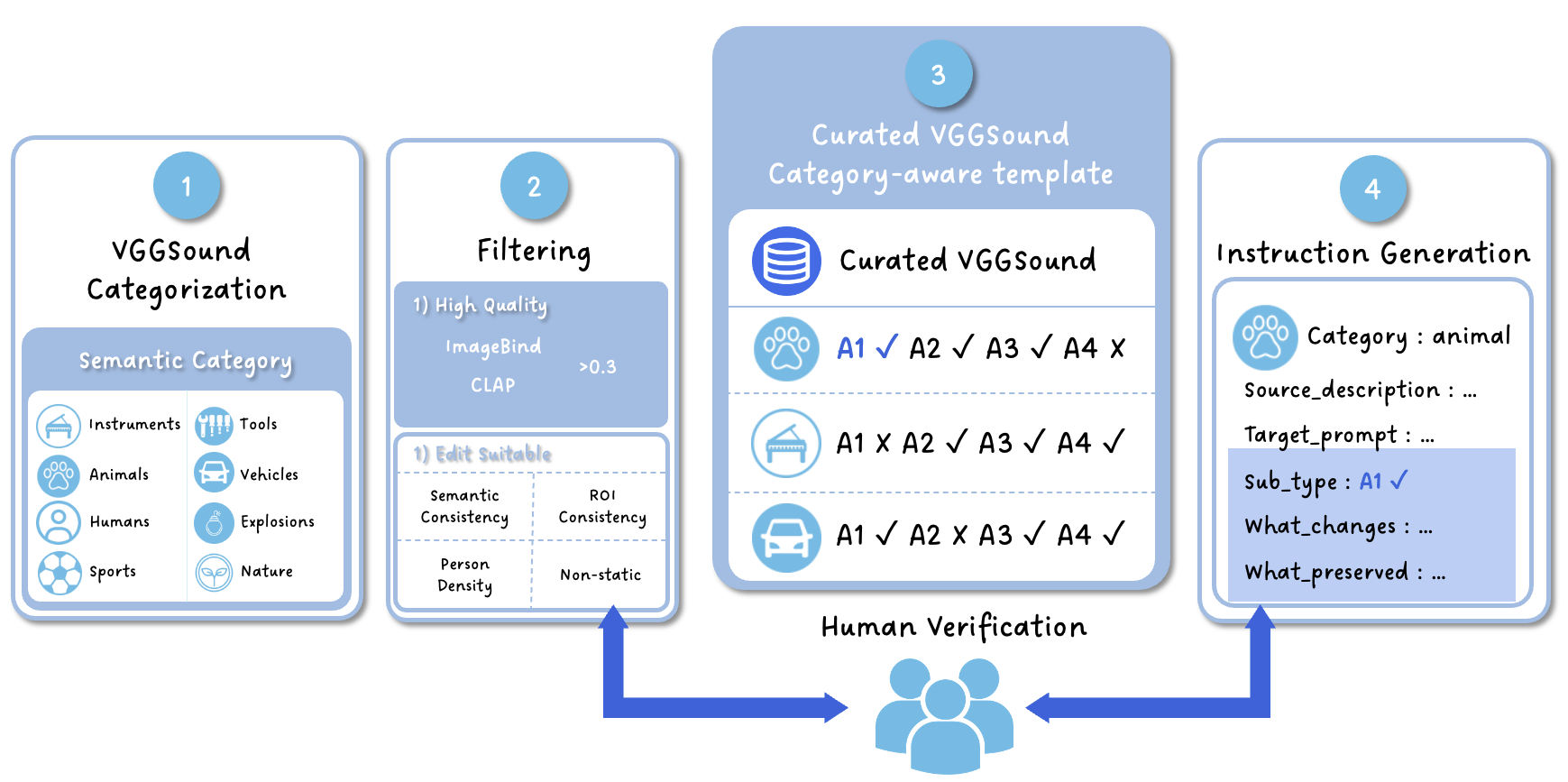}
    \caption{\textbf{Overview of \ours~ Data Curation} (1) VGGSound clips are categorized into 8 semantic categories.
    (2) A two-stage filtering pipeline retains high-quality, edit-suitable clips based on audio-visual alignment and edit compatibility criteria (semantic consistency, ROI consistency, person density, non-static). (3) Human annotators verify the retained clips and confirm which edit types are applicable per clip, producing a curated set with category-aware edit type assignments. (4) Valid clips are paired with structured instruction templates to generate per-sample annotations including target prompt, edit sub-type, what changed and what preserved, which are again verified by human annotators.}
    \label{fig:pipeline}
\end{figure}

In this section, we describe the rigorous multi-stage filtering pipeline employed to construct a high-quality and diverse audio-video editing dataset, as mentioned in the main \Sref{sec:curation}.
For clarity, we distinguish two levels of terminology: our overall
\emph{three-stage curation pipeline} consists of (1) categorization,
(2) filtering, and (3) category-aware instruction generation, whereas the
\emph{two-stage filtering} refers to the two quality filters applied
within stage (2) — audio-video quality filtering and editing-suitability
filtering. Human verification is not an additional filtering stage; as
illustrated in \Fref{fig:pipeline}, it is applied at two separate points
of the pipeline (\Sref{sec:human_verification}).

\subsubsection{Audio-Video Quality Filtering.}
 To ensure that each clip contains well-aligned audio-video content, we apply two complementary audio-video correspondence metrics. We compute audio-video similarity using \textbf{ImageBind}~\cite{girdhar2023imagebind} and audio-language alignment using \textbf{CLAP}~\cite{clap}, retaining only clips for which both scores exceed a threshold of 0.3. This step removes clips where the audio and video streams are poorly correlated, which would undermine the validity of AV-Coupled edit evaluation.
\subsubsection{Editing Suitable Filtering.}
We apply four filters to ensure that retained clips are suitable for video editing tasks.

\paragraph{Semantic Consistency Filter.} We compute pairwise CLIP \cite{radford2021learning}~similarity across 10 uniformly sampled frames (one per second) and discard clips whose mean inter-frame similarity falls below 0.75, excluding clips composed of multiple spliced scenes.

\paragraph{Person Density Filter.} To avoid complex occlusion and identity ambiguity, we apply YOLOv8n~\cite{yolov8_ultralytics} on the 10 sampled frames and reject a clip if $\geq$3 persons are detected in $\geq$5 out of 10 frames.

\paragraph{ROI Consistency Filter.} To ensure a coherent primary subject throughout the clip, we require that the most frequently detected object class appears in at least 9 out of 10 sampled frames. Clips failing this criterion are discarded.

\paragraph{Static Video Filter.} To remove near-static clips that lack meaningful temporal dynamics, we compute the mean SSIM between consecutive sampled frame pairs and discard clips whose average SSIM exceeds 0.85. This threshold empirically captures clips where frames are near-identical throughout, which are unsuitable for video editing tasks. Additionally, for the high-quality test set used in model evaluation, all remaining clips were manually verified to ensure the complete absence of static content.

\subsection{Human Verification}
\label{sec:human_verification}

\begin{figure}[t!]
    \centering
    \includegraphics[width=\linewidth]{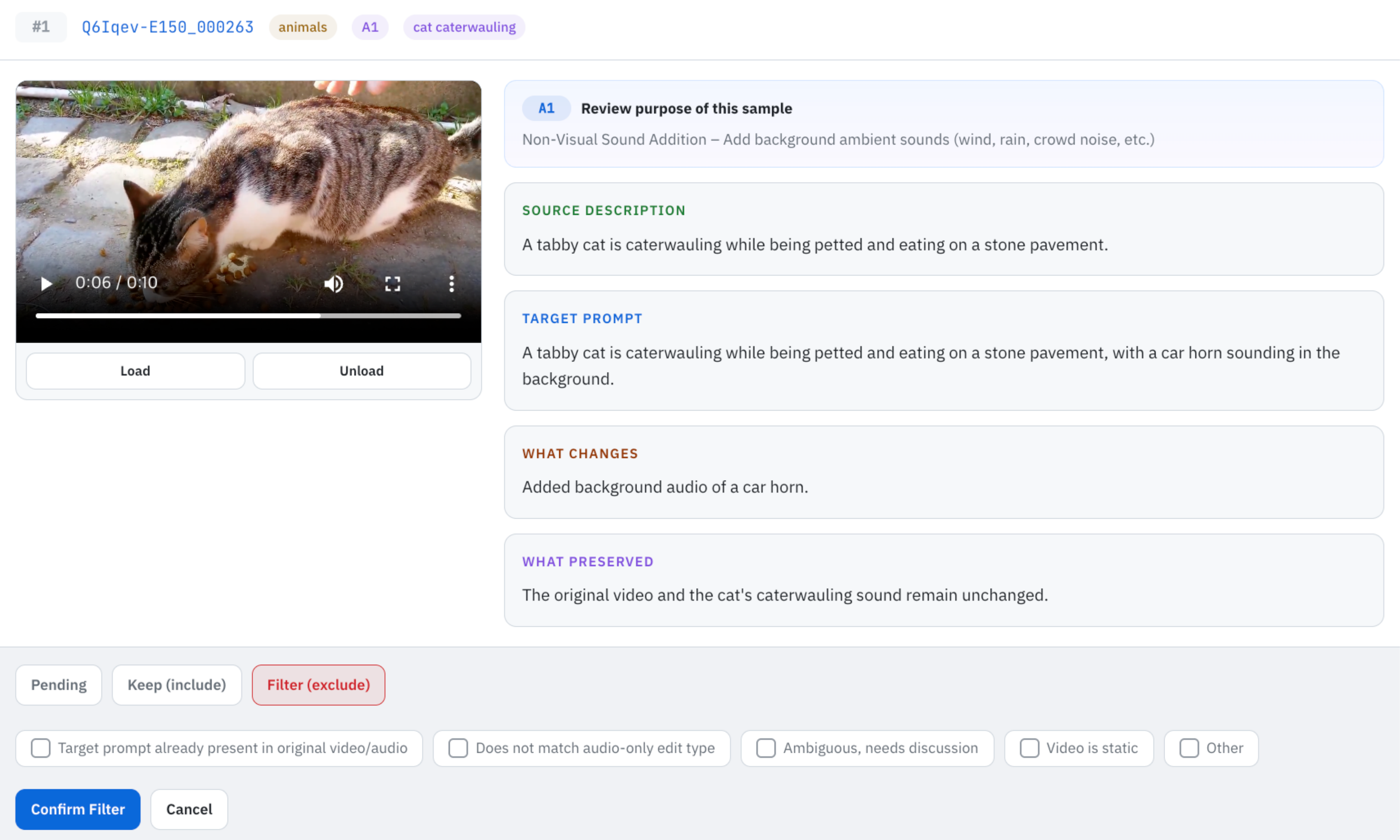}
    \caption{\textbf{Illustration of the user interface used for human verification.}}
    \label{fig:human_verification}
\end{figure}

Human verification is applied at two points of the curation pipeline rather than as a separate filtering stage. First, after the filtering stage, annotators inspect the retained clips to confirm that they are free of residual artifacts such as static content. Second, after instruction generation, annotators review each generated annotation and regenerate inappropriate ones. This section details the latter, and presents an example of the user interface employed during this stage. Two expert annotators participated in verifying the edit instructions generated by the MLLM. Each annotator was presented with the MLLM-generated annotation for the source video — comprising the source prompt, target prompt, what\_changed, and what\_preserved fields — and reviewed all four fields jointly, and flagged an edit instruction for the filtering pool if it satisfied any of the following criteria:

\begin{itemize}
\item The target prompt generated by the MLLM already exists in the source video or audio.
\item The target prompt is designated as an audio-targeted edit, yet its modification would affect other modalities.
\item The target prompt is semantically ambiguous.
\item The source video is a static (i.e., no-motion) video.
\item Other reasons, including cases where the what\_changed or what\_preserved description does not correctly reflect the intended edit.
Note that the listed criteria correspond to commonly observed failure
modes rather than an exhaustive taxonomy; any other issue identified by the
annotators was flagged under "other reasons."

\end{itemize}
Among the flagged samples, annotators manually regenerated appropriate edit instructions for all entries except those involving static videos, which were discarded entirely. \Fref{fig:human_verification}~ illustrates an example of the human verification user interface.

To assess the consistency of the human verification process, we measured inter-annotator agreement on a subset of 180 instances sampled from the full dataset. To ensure coverage across the editing taxonomy, we randomly selected 15 instances from each taxonomy category. The two annotators independently evaluated the same subset using the verification criteria described above.

We report three complementary agreement statistics: raw agreement, Cohen’s $\kappa$~\cite{cohenskappa}, and Gwet’s AC1~\cite{gwet}. Raw agreement measures the proportion of instances for which the two annotators reached the same verification decision, while Cohen’s $\kappa$ accounts for agreement expected by chance. We additionally report Gwet’s AC1, which is less sensitive to prevalence-related instability in chance-corrected agreement measures. The two annotators achieved 90.6\% raw agreement, Cohen’s $\kappa = 0.690$ (95\% CI: 0.540--0.818), and Gwet’s AC1 $= 0.865$. These results indicate a high level of consistency between annotators under the verification protocol.

\begin{table*}[t!]
\centering
\caption{\textbf{Category-aware editing taxonomy.} \checkmark indicates that the edit type is applicable to the given category. Edit types are grouped by modality scope: A (Audio-targeted), B (Video-targeted), and C (AV-coupled).}
\label{tab:taxonomy}
\resizebox{\textwidth}{!}{%
\begin{tabular}{llcccccccc}
\toprule
\textbf{Group} & \textbf{Edit Type} &
  \textbf{Instruments} &
  \textbf{Animals} &
  \textbf{Human} &
  \textbf{Vehicles} &
  \textbf{Domestic} &
  \textbf{Sports} &
  \textbf{Nature} &
  \textbf{Explosions} \\
\midrule
\multirow{4}{*}{\textit{A. Audio-targeted}}
 & A1: Non-visual Sound Addition    & --         & \checkmark & --         & --         & \checkmark & \checkmark & \checkmark & -- \\
 & A2: Non-visual Sound Removal     & \checkmark & \checkmark & --         & \checkmark & \checkmark & \checkmark & \checkmark & -- \\
 & A3: Non-visual Sound Replacement & \checkmark & \checkmark & --         & --         & --         & --         & \checkmark & -- \\
 & A4: Sound Intensity/Pitch Change & \checkmark & --         & --         & \checkmark & --         & --         & \checkmark & \checkmark \\
\midrule
\multirow{3}{*}{\textit{B. Video-targeted}}
 & B1: Attribute/Color Change       & \checkmark & \checkmark & \checkmark & \checkmark & \checkmark & \checkmark & --         & -- \\
 & B2: Style Transfer               & \checkmark & \checkmark & \checkmark & \checkmark & --         & \checkmark & \checkmark & -- \\
 & B3: Background Object Edit       & --         & \checkmark & \checkmark & --         & \checkmark & --         & --         & -- \\
\midrule
\multirow{6}{*}{\textit{C. AV-coupled}}
 & C1: Scene/Weather/Time Change    & --         & \checkmark & --         & \checkmark & --         & --         & \checkmark & -- \\
 & C2: Action Change                & --         & --         & \checkmark & --         & --         & \checkmark & --         & -- \\
 & C3: Sounding Object Swap         & \checkmark & \checkmark & --         & \checkmark & \checkmark & --         & --         & \checkmark \\
 & C4: Mood Change                  & --         & \checkmark & --         & --         & --         & --         & \checkmark & -- \\
 & C5: Speed/Rate Change            & \checkmark & --         & \checkmark & \checkmark & \checkmark & --         & --         & \checkmark \\
\bottomrule
\end{tabular}%
}
\end{table*}

\subsection{Human alignment with automated MLLM Judge}
\label{sec:human_alignment_judge}

To further assess the reliability of our automated evaluation, we conduct a human--judge alignment study that directly compares the preferences of automated MLLM judges with those of human evaluators. We evaluate our primary judge, Qwen3-Omni, together with four alternative omnimodal judges under the same protocol. Below, we describe the human annotation protocol and agreement metrics in detail.

\begin{figure}[t!]
    \centering
    \includegraphics[width=0.9\columnwidth]{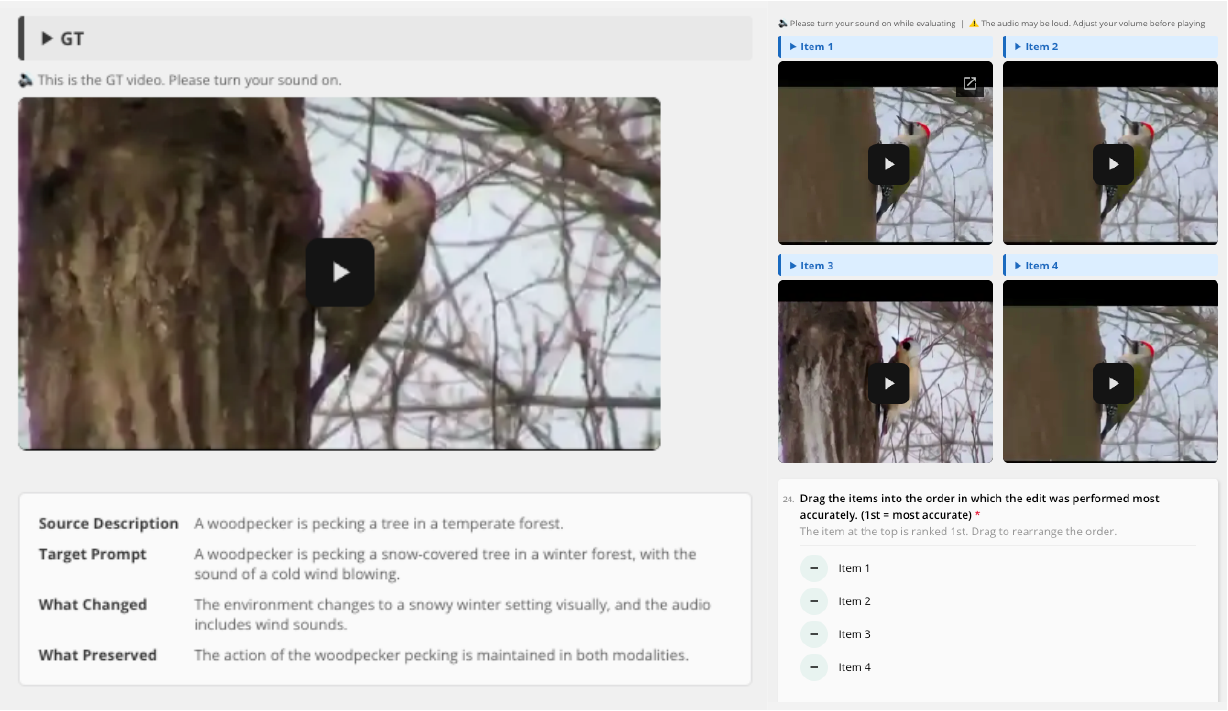}
   \caption{\textbf{User interface for the human--judge alignment study.}}
    \label{fig:human_alignment_study}
\end{figure}

\paragraph{Experimental setup.}
We randomly sample 60 edit instances from \ours. For each instance, human evaluators compare the outputs of the available AV editing systems, resulting in 260 instance-level system pairs in total. Thirteen evaluators independently assess the model outputs according to the same evaluation dimensions used in our automated evaluation, including Edit Accuracy, Modality Selectivity, and AV Consistency. Each evaluator provides a ranking of the candidate outputs for each assigned instance. Figure~\ref{fig:human_alignment_study} shows an example for Edit Accuracy in \cat C. Because human evaluators provide rankings whereas automated judges produce numerical scores, we convert both into pairwise preferences over system outputs. For a pair of systems $(A,B)$, a human preference is obtained directly from the evaluator's relative ranking. For an automated judge, a non-tied score directly determines the preferred system. When the two outputs receive tied scores, we additionally perform a direct pairwise comparison between them. To account for presentation-order bias, this comparison is conducted in both $A$--$B$ and $B$--$A$ orders. We retain only cases that yield the same non-tied preference in both orders; tied or order-inconsistent outcomes are treated as undecided and excluded from the agreement computation.

\paragraph{Automated judges.}
We evaluate five automated judges: Qwen3-Omni~\cite{Qwen3-Omni}, Gemini-3.1-Pro~\cite{gemini31pro}, Gemini-3.6-Flash~\cite{gemini36flash}, Qwen2.5-Omni~\cite{Qwen2.5-Omni}, and AV-Flamingo~\cite{avflamingo}. All judges are evaluated on the same sampled instances and with the same evaluation prompts and pairwise conversion procedure. Qwen3-Omni serves as the primary judge used in our main benchmark evaluation, while the remaining models are included to assess the robustness of the observed human alignment to the choice of judge.

\paragraph{Agreement metrics.}
We report three complementary metrics: individual-level pairwise agreement, leave-one-out agreement, and Kendall's $\tau_b$.

\textbf{Individual-level pairwise agreement.}
For each automated judge, we compute its pairwise agreement separately with each of the 13 human evaluators and average the resulting agreement rates. As the human--human reference, we apply the same pairwise comparison between human evaluators on the exact subset of system pairs resolved by the corresponding judge. We then average the human reference across evaluators. The value reported in the main table is the mean human reference across the five automated judges.

\textbf{Leave-one-out agreement.}
We additionally measure agreement with the human consensus. We hold out one evaluator at a time and construct a reference preference using the majority vote of the remaining 12 evaluators. Both the held-out evaluator and each automated judge are compared against this reference on the judge-specific valid pairs. The process is repeated for all 13 evaluators and the resulting agreement scores are averaged.

\textbf{Kendall's $\tau_b$.}
To evaluate ranking-level consistency, we convert each automated judge's pairwise preferences within an instance into a system ranking using Copeland scores. We compare this ranking with the aggregated human ranking obtained using Borda aggregation and compute Kendall's $\tau_b$, which accounts for ties. For the human reference, each evaluator is compared with the aggregate ranking of the remaining 12 evaluators.

\section{Additional Experiments}
\label{supsec:additional_experiments}
\subsection{Self-Prompt Preservation under Embedding Metrics}
\label{sec:self_prompt_embeddings}
To complement our MLLM-based evaluation, we further assess self-prompt preservation using conventional embedding-based metrics (Table~\ref{tab:self_aug_embedding}). Under the null-edit condition, \textbf{Separated} models exhibit audio deviations (e.g., RAVE+SDEdit, CLAP: 0.721), which reflect instability in the audio editing branch itself, as their audio and video branches operate independently. \textbf{Sequential} models show similar audio instability (CLAP $\approx$ 0.73), since their audio branch synthesizes audio from the edited video without explicit anchoring to the original audio. In contrast, \textbf{Joint} models achieve the strongest preservation across both modalities (V-CLIP: 0.995, LPIPS: 0.110, CLAP: 0.886). Overall, the embedding-based evaluation exhibits trends consistent with our MLLM-based preservation results, providing complementary evidence for the relative preservation behavior of the three editing paradigms.

\begin{table*}[t]
\caption{\textbf{Self-prompt preservation evaluation across AV editing model types.} We evaluate source preservation when the target prompt is identical to the source prompt and therefore requires no semantic edit. Higher V-CLIP~\cite{wang2024videoclip}, ImageBind~\cite{girdhar2023imagebind}, DINO~\cite{oquab2023dinov2}, and CLAP~\cite{clap} scores and lower LPIPS~\cite{lpips} and LPAPS scores indicate stronger preservation. The RAVE + SDEdit, RAVE + ZETA, and RAVE → CAVE pipelines share the same RAVE-generated video output; therefore, their video-preservation scores are identical and are reported once using merged cells spanning the corresponding rows. Audio-preservation scores are evaluated separately for each complete pipeline. For Source → CAVE, the original source video is directly retained without video editing, and "–" therefore denotes that video-preservation metrics are not separately evaluated. \textbf{Bold} and \underline{underlined} values indicate the best and second-best results, respectively.}
\label{tab:self_aug_embedding}
\footnotesize
\centering

\begin{tabular}{ccc|cccc|ccc}
\toprule

\multirow{2}{*}{\textbf{Model Type}} & \multicolumn{2}{c|}{\textbf{Model}} & \multicolumn{4}{c|}{\textbf{Video Preservation}} & \multicolumn{3}{c}{\textbf{Audio Preservation}} \\
\cmidrule(lr){2-3} \cmidrule{4-7} \cmidrule{8-10} & \textbf{Video} & \textbf{Audio} 
  & \textbf{V-CLIP}$\uparrow$ & \textbf{IB}$\uparrow$ 
  & \textbf{DINO}$\uparrow$ & \textbf{LPIPS}$\downarrow$
  & \textbf{CLAP}$\uparrow$ & \textbf{IB}$\uparrow$ 
  & \textbf{LPAPS}$\downarrow$ \\
\midrule
 \multirow{2}{*}{\textit{Separated}} & \multicolumn{2}{c|}{RAVE $\quad+\quad$ SDEdit} 
  & \multirow{3}{*}{\underline{0.956}} & \multirow{3}{*}{\underline{0.911}} 
  & \multirow{3}{*}{\underline{0.939}} & \multirow{3}{*}{\underline{0.204}} 
  & 0.721 & 0.534 & 4.111 \\

& \multicolumn{2}{c|}{RAVE$\quad+\quad$ZETA} 
  &  &  &  & 
  & \underline{0.859} & \underline{0.757} & \textbf{2.765} \\
\cmidrule{1-3} \cmidrule{8-10}  

\multirow{2}{*}{\textit{Sequential}} & \multicolumn{2}{c|}{RAVE $\quad\rightarrow\quad$ CAVE}        
  &  &  &  & 
  & 0.736 & 0.589 & 4.323 \\
& \multicolumn{2}{c|}{Source $v$$\quad\rightarrow\quad$CAVE}    
   & \cellcolor{gray!40}- & \cellcolor{gray!40}- & \cellcolor{gray!40}- & \cellcolor{gray!40}-
  & 0.731 & 0.580 & 4.320 \\
\midrule

\textit{Joint} & \multicolumn{2}{c|}{AvED}       
  & \textbf{0.995} & \textbf{0.939} & \textbf{0.967} & \textbf{0.110} 
  & \textbf{0.886} & \textbf{0.846} & \underline{3.126} \\
\bottomrule
\end{tabular}
\end{table*}

\subsection{Cross-Modality-Conditioned Targeted Editing}
\label{supsec:cross_modal_cond}

Our main benchmark deliberately isolates single-modality edits so that modality-selective preservation can be measured cleanly. A complementary setting, however, is one in which a single-modality edit must be \emph{anchored to an event in the other modality}---for example, \textit{"add a car horn when the cat yawns."} Such prompts require not only producing the correct edit content, but also localizing it in time with respect to a trigger observed in the untargeted modality. We therefore construct a pilot study to examine whether current AV editing models are capable of this form of cross-modal reasoning.

\paragraph{Pilot setup.}
We build a pilot set of 70 samples (40 audio-targeted and 30 video-targeted), each manually curated and verified by human annotators, in which every single-modality edit is conditioned on a trigger event in the other modality. 
To evaluate this setting we introduce an additional axis, \textbf{Cross-Modal Temporal Grounding (CMTG)}, assessed with the same MLLM-as-a-judge framework described in \Sref{sec:evaluation}. CMTG is reported separately from Edit Accuracy (EA) because the two ask different questions: EA asks whether the correct edit content was produced, whereas CMTG asks whether that edit is temporally grounded on the trigger---e.g., whether the horn onsets at the yawn rather than sounding throughout the clip. We compare a joint model (AvED) and a separated pipeline (RAVE\,$+$\,ZETA); MMAudio is excluded for the same reason as in \Sref{sec:self_aug}. All scores are scaled to 100, with higher values indicating better performance.

\paragraph{Results.}
\Tref{tab:cross_modal_cond} reports the results. Two observations stand out. 
First, introducing a cross-modal condition does not degrade edit content: EA remains comparable to, and in most cases slightly above, the non-conditioned setting on the same clips. Yet CMTG is uniformly low across all models (20--29 out of 100). Current models thus still produce the right edit \emph{content} while largely failing to align it with the trigger event in the other modality, indicating that cross-modal temporal grounding remains a genuinely open capability rather than an incidental weakness.
Second, the ranking between paradigms is not preserved across the two axes. The separated pipeline is strongest on audio-targeted EA (87.0), but this advantage disappears on CMTG (24.6 vs.\ 28.6 for the joint model). This suggests that the apparent strength of separated pipelines is largely confined to settings with little cross-modal dependency, and becomes less pronounced once an edit must be conditioned on the other modality. We note, however, that the joint model's edge should not be read as a solution: AvED localizes edits via cross-attention masks and a contrastive objective over prompt-relevant regions---a mechanism not designed for temporal grounding---so its margin on CMTG reflects being \emph{less poor} rather than competent. No paradigm resolves this setting.

\paragraph{Discussion.}
These results indicate that cross-modality-conditioned editing is a meaningful and currently underexplored axis that fits naturally alongside our existing categories: cleanly separated edits capture modality-selective preservation, while conditioned edits additionally probe cross-modal reasoning, and the two together give a fuller picture of AV editing capability. Scaling this pilot into a dedicated cross-modal-conditioned track within the audio- and video-targeted categories, and reporting CMTG alongside our existing metrics, is a direct extension of \ours.

\begin{table}[t]
\centering
\footnotesize
\caption{\textbf{Cross-modality-conditioned targeted editing (pilot study).}
Each edit is anchored to a trigger event in the untargeted modality.
\textit{EA (non-cond.)} reports Edit Accuracy on the original, non-conditioned
prompts of the same clips as a reference point, while \textit{EA} is measured
on the cross-modal-conditioned prompts. \textit{CMTG} denotes Cross-Modal
Temporal Grounding. All scores are scaled to 100; higher is better.
\textbf{Bold} indicates the better result between the two paradigms.}
\vspace{2mm}
\label{tab:cross_modal_cond}
\begin{tabular}{ll|cc}
\toprule
\textbf{Category} & \textbf{Metric} & \textbf{Joint} & \textbf{Separated} \\
\midrule
\multirow{3}{*}{\textit{Audio-targeted}}
  & EA (non-cond.)$\uparrow$ & 63.4 & \textbf{81.6} \\
  & EA$\uparrow$             & 70.0 & \textbf{87.0} \\
  & CMTG$\uparrow$           & \textbf{28.6} & 24.6 \\
\midrule
\multirow{3}{*}{\textit{Video-targeted}}
  & EA (non-cond.)$\uparrow$ & 72.6 & \textbf{77.4} \\
  & EA$\uparrow$             & \textbf{74.0} & 68.0 \\
  & CMTG$\uparrow$           & \textbf{22.0} & 20.0 \\
\bottomrule
\end{tabular}
\end{table}

\subsection{Consistency with Embedding-based Evaluation}
\label{supsec:embedding_based_eval}
\begin{figure}[t!]
    \centering
    \includegraphics[width=\linewidth]{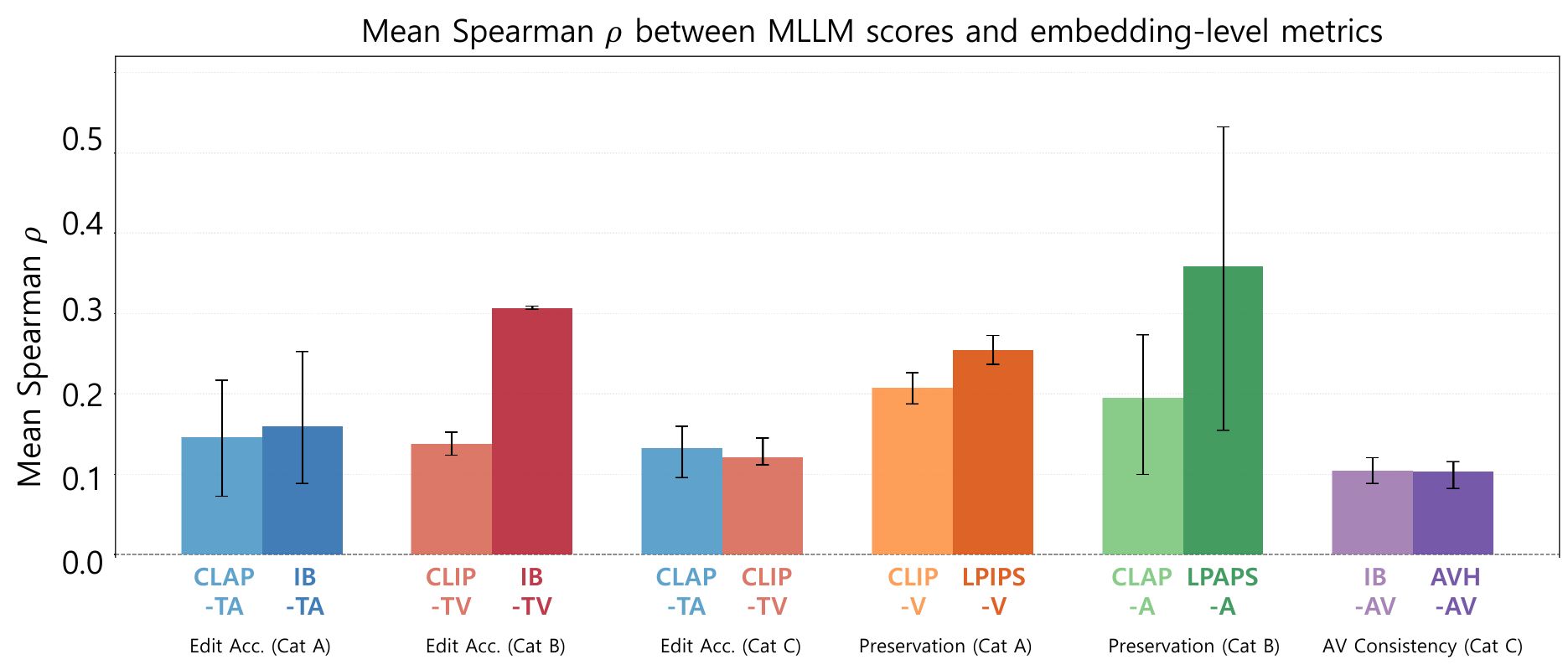}
    \caption{
    \textbf{Mean Spearman $\rho$ between our MLLM-based evaluation scores and embedding based metrics across all evaluation dimensions and edit categories.} Error bars indicate the min–max range across generation models. All reported correlations are statistically significant $(p < 0.05)$. Higher $\rho$ indicates stronger agreement between MLLM-based scores and embedding-based metrics.  
The suffixes \texttt{ta}, \texttt{tv}, and \texttt{av} denote target text--target audio, target text--target video, and target audio--target video alignment, respectively. 
The single-modality suffixes \texttt{a} and \texttt{v} instead denote preservation metrics, measuring source audio--target audio and source video--target video alignment, respectively.}
    \label{fig:spearman}
\end{figure}

To validate our MLLM-based evaluation, we compute the Spearman rank correlation ($\rho$) between MLLM scores and representative mid-level embedding metrics for each evaluation dimension and edit category. Specifically, we report the mean $\rho$ averaged across all evaluated AV editing models. In Figure~\ref{fig:spearman}, AVHScore~\cite{mao2024tavgbench} quantifies audio-visual semantic consistency by averaging the cosine similarity between each video-frame embedding and the audio embedding in a shared embedding space. Higher values indicate that the generated audio and video are more semantically aligned. Suffixes denote the modalities used in each metric. As shown in \Fref{fig:spearman}, all 12 correlations are positive and statistically significant ($p<0.05$), confirming that our MLLM-based scores consistently reflect the quality signal captured by embedding-based metrics. Modality Selectivity yields the strongest $\rho$ across categories, indicating that MLLM judgments on audio and video preservation closely align with acoustic and visual similarity measures. Edit Accuracy in \cat B also shows relatively strong $\rho$, while AV Consistency correlations are modest but consistent across all categories. These results support the reliability of our MLLM-based evaluation as a holistic assessment tool that extends beyond individual embedding metrics.

\subsection{Overall scores of AV editing models for each sub-category \cat~C}
\begin{figure}[t]
    \centering
    \includegraphics[width=\linewidth]{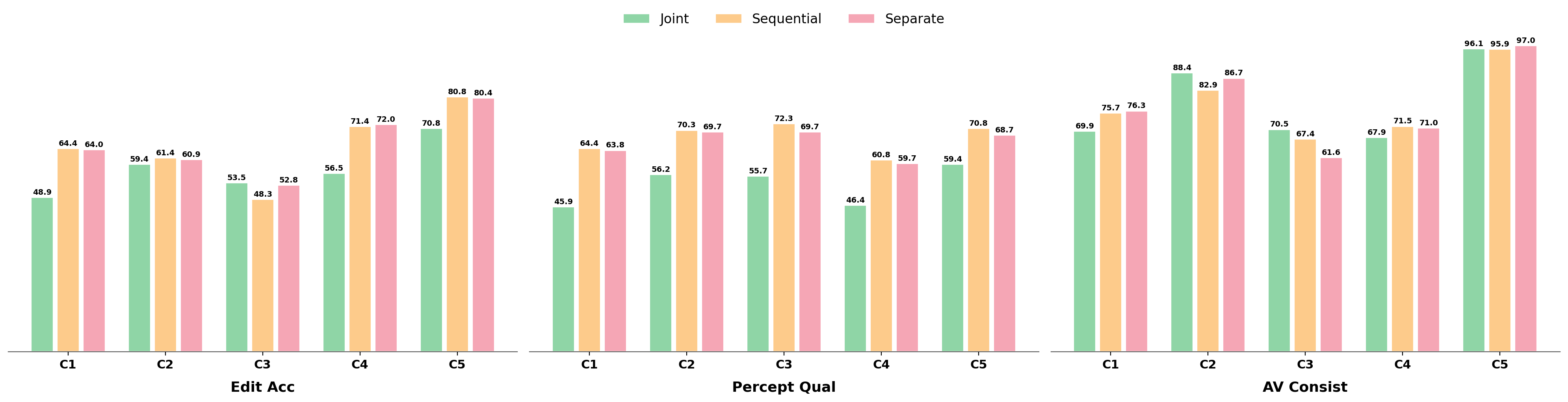}
    \caption{\textbf{Category C overall performance}}
    \label{fig:cat_C}
\end{figure}

Figure~\ref{fig:cat_C} visualizes the overall scores of AV editing models for each sub-category of \cat~C.

\subsection{Audio-conditioned Video Editing with AVIEdit}

To extend our evaluation to audio-conditioned video editing, we additionally evaluate AVIEdit~\cite{aviedit} on \ours. AVIEdit requires both an edited audio track and a spatial mask specifying the video region to be modified. To construct these inputs consistently across the benchmark, we first use Qwen3~\cite{qwen3llm} to extract the target object or region from each edit instruction and then localize the corresponding region using Grounded SAM-2~\cite{groundedsam}. The resulting mask, together with the audio edited by ZETA, is provided to AVIEdit. When Grounded SAM-2 fails to return a valid region, we use a full-frame mask as a fallback.

\paragraph{AV-coupled editing.}
Table~\ref{tab:av_coupled_edit_accuracy} reports model-wise Edit Accuracy for each AV-coupled sub-category. AVIEdit performs relatively weakly on global transformations such as Scene Change (C1) and Mood Change (C4), achieving 39.2 in both categories. This is consistent with its reliance on spatially localized masks, which are less suited to edits requiring global scene-level modification. In contrast, AVIEdit achieves performance comparable to the other evaluated models on Action Change (C2), Object Replacement (C3), and Speed Change (C5), with Edit Accuracy scores of 60.4, 50.5, and 77.0, respectively. These results suggest that audio-conditioned localized video editing is more effective when the requested modification can be associated with a specific object, action, or spatial region.

\paragraph{Video-side evaluation.}
Table~\ref{tab:video_modality_evaluation} further evaluates the sequential ZETA$\rightarrow$AVIEdit pipeline using the same video-side protocol as Table~\ref{tab:video_modality_evaluation} in the main paper. The pipeline achieves the highest Perceptual Quality for Video-targeted editing (73.9), exceeding RAVE+ZETA (69.5) and AvED (53.2). However, its Edit Accuracy is lower (65.0) than both RAVE+ZETA (79.1) and AvED (77.3). Its video Modality Selectivity under Audio-targeted editing is also slightly lower (84.6) than RAVE+ZETA (86.2) and AvED (87.5), indicating that audio-conditioned video editing can introduce unintended changes to the visual stream even when the video should be preserved.

Overall, AVIEdit produces perceptually plausible video outputs but does not consistently improve instruction fidelity or preservation of non-target visual content. Its inclusion broadens our evaluation to an explicit audio-conditioned video editing pipeline and further highlights the trade-off among edit accuracy, perceptual quality, and preservation in current AV editing systems.

\begin{table}[t!]
\centering
\footnotesize
\caption{\textbf{AV-coupled (C) Edit Accuracy.} \textbf{Bold} denotes the best result in each column.}
\vspace{2mm}
\label{tab:av_coupled_edit_accuracy}
\begin{tabular}{lccccc}
\toprule
\textbf{Model} & \textbf{\shortstack{Scene\\change\\(C1)}} & \textbf{\shortstack{Action\\change\\(C2) $\uparrow$}}
  & \textbf{\shortstack{Object\\replacement\\(C3) $\uparrow$}} & \textbf{\shortstack{Mood\\change\\(C4) $\uparrow$}} & \textbf{\shortstack{Speed\\change\\(C5) $\uparrow$}} \\
\midrule
RAVE$\quad+\quad$ZETA   & \textbf{64.6} & 59.7          & 51.5          & 70.2          & 77.9 \\
RAVE$\quad+\quad$SDEdit & 63.4          & \textbf{62.1} & \textbf{54.1} & \textbf{73.9} & \textbf{82.9} \\
CAVE        & \underline{64.4}          & \underline{61.4}          & 48.3          & \underline{71.4}          & \underline{80.8} \\
AVIEdit     & 39.2          & 60.4          & 50.5          & 39.2          & 77.0 \\
AvED        & 48.9          & 59.4          & \underline{53.5}          & 56.5          & 70.8 \\
\bottomrule
\end{tabular}
\end{table}

\begin{table*}[t!]
\small
\centering
\caption{\textbf{Video modality evaluation overview.} All scores are scaled to 100. Edit Accuracy and Perceptual Quality are measured on Cat.B (video-targeted) samples, and Modality Selectivity on Cat.A (audio-targeted) samples to assess unintended video change. Higher is better for all metrics.}
\label{tab:video_modality_evaluation}
\begin{tabular}{l c@{\,}c@{\,}c ccc}
\toprule
\multirow{2}{*}{\textbf{Model Type}} & \multicolumn{3}{c}{\textbf{Model}} & \textbf{Edit Accuracy} & \textbf{Percept Quality} & \textbf{Modal Selectivity~(Video)} \\
\cmidrule(lr){2-4} \cmidrule(lr){5-5} \cmidrule(lr){6-6} \cmidrule(lr){7-7}
& \textbf{Video} & & \textbf{Audio} & (Cat B)$\uparrow$ & (Cat B)$\uparrow$ & (Cat A)$\uparrow$ \\
\midrule
\textit{Separate}   & RAVE & $+$          & ZETA    & \textbf{79.1} & \underline{69.5}          & \underline{86.2} \\
\midrule
\textit{Sequential} & ZETA & $\rightarrow$ & AVIEdit & 65.0          & \textbf{73.9} & 84.6 \\
\midrule
\textit{Joint} & \multicolumn{3}{c}{AvED} & \underline{77.3} & 53.2 & \textbf{87.5} \\
\bottomrule
\end{tabular}
\end{table*}


\definecolor{bestcell}{RGB}{255, 243, 205}  

\begin{table*}[h]
  \footnotesize
  \centering                                                                    
  \caption{\textbf{Existing AV editing models evaluated across our proposed edit taxonomies.} $^{*}$Video editing via shared RAVE~\cite{rave} backbone.}
  \label{tab:models}
  \begin{tabular}{llccc}
  
  \toprule
  \textbf{Model} & \textbf{Paradigm} & \textbf{A(Audio)} & \textbf{B(Video)} &      
  \textbf{C(AV)} \\                                                        
  \midrule
  \aved~\cite{aved}         & Joint      & \checkmark & \checkmark        &      
  \checkmark \\   
  \midrule
  \cave~\cite{coherent}     & Sequential & \checkmark & \checkmark$^{*}$  &      
  \checkmark \\
  MMAudio~\cite{mmaudio}   & Sequential & \checkmark & \checkmark$^{*}$  &      
  \checkmark \\                                                                 
  \midrule
  RAVE$\quad+\quad$SDEdit~\cite{rave,sdedit} & Separate & \checkmark & \checkmark$^{*}$ &
  \checkmark \\    
  RAVE$\quad+\quad$ZETA~\cite{rave,zeta}     & Separate & \checkmark & \checkmark$^{*}$ &   
  \checkmark \\
  \bottomrule
  \end{tabular}                                                                 
  \end{table*}

\section{Implementation Details}
\label{sec:implementation}

\paragraph{Baselines}
We evaluate the models listed in \Tref{tab:models}, running each under its proposed default configuration without modification. For \textbf{AvED}, since its temporal shuffling operates on a $2{\times}2$ image grid, the frame rate is fixed at 4 fps. For \textbf{RAVE}, the fps is adjusted to match the original clip length, ensuring the output is generated at the source fps. For \textbf{MMAudio (CAVE)}, since MMAudio is trained on 8-second clips, inputs are automatically truncated to 8 seconds upon entry.

All models are constrained to produce outputs at the original source fps. For fair evaluation, when comparing results across models on the same sample, all outputs are trimmed to the shortest duration among all models for that sample.

Regarding audio sampling rates, \textbf{ZETA} and \textbf{SDEdit} are based on AudioLDM2 and thus operate at 16,000 Hz. \textbf{MMAudio} uses the \texttt{small\_16k} variant to match this rate. \textbf{CoherentAVEdit} provides a fine-tuned model built on top of MMAudio, which also operates at 16,000 Hz.

\paragraph{Computational Cost}
All models are evaluated on a single NVIDIA A6000 (48GB) GPU with a batch size of 1. Audio-based models — \textbf{CoherentAVEdit}, \textbf{MMAudio}, \textbf{ZETA}, and \textbf{SDEdit} — complete inference within 1 minute per sample. \textbf{AvED} requires approximately 20 minutes per sample due to its $2{\times}2$ grid-based temporal shuffling mechanism, while \textbf{RAVE} takes approximately 35 minutes per sample.

\paragraph{Experimental Setup} To comprehensively assess how each editing paradigm handles the \textit{audio modality}, we evaluate models using samples from two benchmark categories. First, using \cat A (audio-targeted editing) samples, we measure edit accuracy and perceptual quality to quantify how faithfully and naturally each model performs the intended audio edit. Second, using \cat B (video editing) samples, where only the video is edited, we measure modality selectivity to assess whether the audio is preserved intact—a critical requirement for practical editing. For sequential paradigms, where the audio model takes the video model's output as input, we additionally include an source $v$ condition that bypasses the video editing stage. This ablation isolates the audio model's inherent capability from any artifacts propagated through the sequential pipeline, revealing whether performance differences stem from the audio model itself or from upstream video editing.

\section{Dataset License}
\label{license}
The VGGSound dataset is released under the Creative Commons Attribution 4.0 International (CC BY 4.0) license. Our use of this dataset complies with its terms, and any modification or redistribution of derived data is subject to the same license conditions. We refer readers to \url{https://creativecommons.org/licenses/by/4.0/} for full license details.

\section{Limitations}
\label{limitation}
While AVENUE enables systematic evaluation and analysis of existing AV editing models across audio-targeted and video-targeted modalities with diverse edit categories — aspects largely overlooked in prior benchmarks — our work has several limitations. First, the edit instructions in our benchmark are generated by a single multimodal LLM (Gemini-3-Flash-preview), which may introduce model-specific biases into the dataset. To mitigate this, each video–prompt pair is manually verified; however, this makes our curation pipeline difficult to scale, and automating the verification stage remains an open challenge. Second, although we apply rigorous multi-stage filtering, the inherent noisiness of the source dataset VGGSound may leave residual artifacts in audio or video quality. Moreover, since all VGGSound clips are restricted to roughly 10 seconds, AVENUE cannot assess model behavior on longer-range temporal coherence or narrative transitions. Extending the benchmark with longer-form AV sources is left for future work. Third, we restrict edit instructions to declarative prompts to support controlled, diagnostic evaluation. However, this design limits coverage of free-form instructions, including imperative commands and other diverse instruction styles. Fourth, our paradigm analysis is limited to sequential, joint, and separated architectures, whereas a broader range of AV editing paradigms may exist. We aim to address these limitations in future work.

\section{Broader Impacts}
\label{broader_impact}
Our \ours~ benchmark provides the audio-visual editing research community with a standardized evaluation framework covering three key editing scenarios: audio-targeted, video-targeted, and AV-coupled editing. By publicly releasing 
the dataset and evaluation code, we aim to facilitate fair and reproducible comparisons across future AV editing models, accelerating progress in the field.

Furthermore, as with any dataset, there may be inherent biases in the scenarios and data included in \ours. We have made efforts to ensure diversity across editing scenarios, but we encourage the community to be mindful of these potential limitations when applying models evaluated on our benchmark to real-world settings.

\section{Ethics Statement}
\label{ethics}
This work constructs the AVENUE dataset from VGGSound, which is publicly available under the CC BY 4.0 license, and uses exclusively the test split to avoid data leakage. No personal data or private information is collected. All editing instructions are synthetically generated via LLMs and do not contain sensitive or harmful content. While our framework is intended solely for research purposes, we acknowledge that audio-visual editing techniques may be misused for generating deceptive content, and encourage responsible use within the research community.

\section{LLM Usage}
\label{llmusage}
In this work, we leverage a multimodal LLM~\cite{Qwen3-Omni} as a judge for evaluation purposes. Beyond this, LLMs were additionally utilized in the following aspects of this research: editing of written content (e.g., grammar, spelling, and word choice), visualizing results for submission, and data processing and filtering.

\section{Prompt Detail}
\subsection{Instruction Generation}
\promptbox{Instruction Generation(Gemini-3-Flash-preview)}{You are an expert in audio-visual editing evaluation. You will be given a video clip and its audio label describing the sound in the scene. \\
Your task: Generate realistic edit prompts for this scene matching with {category}, mapped to the taxonomy categories below. \\
Taxonomy Categories : \{category\_taxonomies\} \\

Instructions \\
1. Analyze the video and audio label to understand the scene. \\
2. For EACH sub-type above, determine whether a plausible edit exists for this specific scene. \\
3. If a plausible edit exists, generate it. If not (e.g., no background object to remove), skip that sub-type. \\
4. Use the exact format below for each edit. \\

\textbf{Output Format} \\
Return a JSON list. Each entry must follow this structure: \\
    "category": "A|B|C", \\
    "sub\_type": "A1|A2|A3|A4|B1|B2|B3|C1|C2|C3|C4|C5", \\
    "source\_description": "<describe the original scene in one sentence>", \\
    "target\_prompt": "<the editing instruction or target description>", \\
    "what\_changes": "<briefly describe what should change>", \\
    "what\_preserved": "<briefly describe what must stay the same>" \\

Example \\
...$($ommision$)$

Important Rules \\
- Generate ONLY edits that are semantically plausible for the given scene. Do not force unnatural edits. \\
- The source\_description should accurately reflect the video and audio label. \\
- The target\_prompt must be a complete scene description (not just the delta), written naturally. \\
- For Audio-Only edits: the target\_prompt should describe the full audio scene. \\
- For Video-Only edits: the target\_prompt should describe the full visual scene. \\
- For AV-Coupled edits: the target\_prompt should describe both modalities. \\

Now, here is your input: \\
Audio Label: \{audio\_label\} \\
Video: \{attached video\} \\
Generate all plausible edit prompts mapped to the taxonomy. \\ }
\label{promtp:inst}
\newpage

\subsection{Category-aware editing taxonomy (category\_taxonomies.txt)}
\label{sup_sec:category_aware_editing_taxonomy}
\promptbox{animal.txt}{\textbf{Category A: Audio-Only Edits} (video must stay unchanged)\\
  A1. Ambient Sound Addition – Add background environmental sounds (e.g., wind, rain, rustling leaves)\\
  A2. Background Sound Removal – Remove background sounds while keeping the animal's sound intact\\
  A3. Background Sound Replacement – Replace background ambient sounds with different environmental sounds\\

\textbf{Category B: Video-Only Edits} (audio must stay unchanged)\\
  B1. Animal Attribute Change – Modify the animal's appearance (e.g., fur color, size, pattern)\\
  B2. Style Transfer – Apply a visual style to the scene (anime, watercolor, sketch, etc.)\\
  B3. Background Vegetation Edit – Add or remove background elements like trees, grass, or bushes\\

\textbf{Category C: AV-Coupled Edits} (both audio and video must change together)\\
  C1. Scene Change – Change the environment or setting (e.g., mountain $\rightarrow$ beach, forest $\rightarrow$ desert)\\
  C3. Animal Swap/Duplicate – Replace the animal with a different species, or duplicate the same animal in the scene\\
  C4. Mood Change – Modify the overall atmosphere or emotional tone of the scene}

\promptbox{domestic\_and\_tools.txt}{\textbf{Category A: Audio-Only Edits} (video must stay unchanged)\\
  A1. Domestic Ambient Addition – Add everyday background sounds (e.g., TV noise, clock ticking, kitchen hum)\\
  A2. Noise Removal – Remove extraneous noise or unwanted background sounds\\

\textbf{Category B: Video-Only Edits} (audio must stay unchanged)\\
  B1. Tool Color Change – Modify the color or appearance of the tool/appliance\\
  B3. Background Object Addition – Add a new object to the background of the scene\\

\textbf{Category C: AV-Coupled Edits} (both audio and video must change together)\\
  C3. Tool Swap/Add/Remove – Add, remove, or replace the tool or appliance with a different one\\
  C6. Machine Speed Change – Modify the operating speed of the machine or tool}

\promptbox{human\_sounds.txt}{\textbf{Category B: Video-Only Edits} (audio must stay unchanged)\\
  B1. Person Attribute Change – Modify clothing color, hairstyle, or other human appearance attributes\\
  B2. Style Transfer – Apply a visual style to the scene (anime, watercolor, sketch, etc.)\\
  B3. Indoor Object Swap – Add, remove, or replace indoor objects/furniture in the background\\

\textbf{Category C: AV-Coupled Edits} (both audio and video must change together)\\
  C2. Action Change – Replace the person's action with a different one (e.g., clapping $\rightarrow$ snapping fingers)\\
  C6. Motion Speed Change – Modify the speed of the person's movement or action (e.g., slow clap $\rightarrow$ fast clap)}

\promptbox{instruments.txt}{\textbf{Category A: Audio-Only Edits} (video must stay unchanged)\\
  A2. Background Instrument Removal – Remove other instruments playing together in the background, isolating the main instrument\\
  A3. Background Instrument Replacement – Replace background ensemble instrument sounds with different instruments\\
  A4. Pitch/Loudness Shift – Change the pitch or loudness of the performance\\

\textbf{Category B: Video-Only Edits} (audio must stay unchanged)\\
  B1. Instrument Color Change – Modify the color or finish of the instrument (e.g., black piano $\rightarrow$ white piano)\\
  B2. Style Transfer – Apply a visual style to the scene (anime, watercolor, sketch, etc.)\\

\textbf{Category C: AV-Coupled Edits} (both audio and video must change together)\\
  C3. Instrument Swap/Add – Remove or replace the instrument with a different one; if the scene is filmed from a distance, add another instrument to the ensemble\\
  C5. Instrument Variant Change – Change the variant or type of the same instrument family (e.g., acoustic guitar $\rightarrow$ electric guitar)\\
  C6. Performance Speed Change – Modify the tempo/speed of the performance (e.g., slow ballad $\rightarrow$ fast virtuoso)}

\promptbox{nature\_and\_environment.txt}{\textbf{Category A: Audio-Only Edits} (video must stay unchanged)\\
  A1. Sound Addition – Add a new natural sound to the scene (e.g., bird calls, insect chirping)\\
  A2. Sound Removal – Remove a specific sound layer from the scene\\
  A3. Sound Replacement – Replace a natural sound with a different one (e.g., seagull calls → crow cawing)\\
  A4. Sound Intensity Change – Adjust the intensity or loudness of a natural sound (e.g., gentle waves → crashing waves)\\

\textbf{Category B: Video-Only Edits} (audio must stay unchanged)\\
  B2. Style Transfer – Apply a visual style to the scene (anime, watercolor, sketch, etc.)\\

\textbf{Category C: AV-Coupled Edits} (both audio and video must change together)\\
  C1. Weather Change – Change the weather or atmospheric conditions (e.g., sunny → rainy, calm → stormy)\\
  C4. Mood Change – Modify the atmosphere to evoke a different emotion (e.g., peaceful forest → eerie forest)}

\promptbox{sports\_and\_leisure.txt}{\textbf{Category A: Audio-Only Edits} (video must stay unchanged)\\
  A1. Crowd Sound Removal – Remove crowd cheering or audience noise from the scene\\
  A2. Crowd Sound Addition – Add crowd cheering, applause, or audience reactions\\

\textbf{Category B: Video-Only Edits} (audio must stay unchanged)\\
  B1. Attribute Change – Modify appearance attributes (e.g., jersey color, equipment color)\\
  B2. Style Transfer – Apply a visual style to the scene (anime, watercolor, sketch, etc.)\\

\textbf{Category C: AV-Coupled Edits} (both audio and video must change together)\\
  C2. Intensity Change – Reduce or increase the intensity of the athletic action (e.g., weaker hit, slower run, softer kick)}

\promptbox{vehicles\_and\_engines.txt}{\textbf{Category A: Audio-Only Edits} (video must stay unchanged)\\
  A2. Engine Sound Removal – Remove or silence the engine/motor sound\\
  A4. Engine Loudness Change – Adjust the loudness of the engine sound (e.g., louder revving, quieter idle)\\

\textbf{Category B: Video-Only Edits} (audio must stay unchanged)\\
  B1. Vehicle Color Change – Modify the color or paint of the vehicle\\
  B2. Style Transfer – Apply a visual style to the scene (anime, watercolor, sketch, etc.)\\

\textbf{Category C: AV-Coupled Edits} (both audio and video must change together)\\
  C1. Time-of-Day Change – Change the lighting/time of the scene (e.g., daytime → nighttime)\\
  C3. Vehicle Type Swap – Replace the vehicle with a different type (e.g., car → police car, sedan → ambulance)\\
  C6. Vehicle Speed Change – Modify the speed of the vehicle (e.g., cruising → racing, slow → fast)}

\promptbox{weapons\_and\_explosions.txt}{\textbf{Category A: Audio-Only Edits} (video must stay unchanged)\\
  A4. Explosion/Gunshot Intensity Change – Adjust the loudness or pitch of explosion or gunfire sounds\\

\textbf{Category C: AV-Coupled Edits} (both audio and video must change together)\\
  C3. Weapon Swap – Replace the weapon with a different type (e.g., machine gun → laser gun, pistol → shotgun)\\
  C6. Firing Rate Change – Modify the speed of firing or detonation (e.g., single shot → rapid fire)}

\newpage

\subsection{MLLM-as-a-Judge prompt}
\label{sec:mllm}
\subsubsection{Edit Accuracy}
\promptbox{Edit Accuracy(EA) - Audio targeted edit}{You are an expert evaluator for audio-video editing quality. Your task is to assess whether an AUDIO edit was performed correctly.
\newline\newline
{[Source audio]}: (attached) \newline
{[Edited audio]}: (attached)
\newline\newline
{[Source description]}: "{source\_description}"
\newline
{[Target description]}: "{target\_prompt}"
\newline\newline
=== EVALUATION CRITERIA (human-verified) === \newline
Changes intended:
  "{what\_changed}" \newline

Elements that must remain intact:
  "{what\_preserved}" \newline
============================================ \newline

Evaluate the edited audio based on the criteria above.
\\
Scoring rubric (1-5): \\
5: All intended changes are fully and precisely realized,
   and all preserved elements remain completely intact.\\
4: Intended changes are mostly realized with minor inaccuracies
   (e.g., correct sound type but slightly off in quality or timing),
   preserved elements are intact.\\
3: Intended changes are partially realized
   (e.g., some changes applied correctly, others missing or weak),
   OR preserved elements show minor unintended damage.\\
2: Intended changes are attempted but mostly incorrect or incomplete,
   OR preserved elements are significantly damaged.\\
1: No meaningful change was made,
   OR changes go in the wrong direction,
   OR preserved elements are destroyed.\\

Provide your response in this format:\\
(1) Change assessment: For each intended change, describe whether
    and how well it was achieved.\\
(2) Preservation assessment: For each preserved element, describe
    whether it remains intact.\\
(3) Score: [1-5]}
\newpage
\promptbox{Edit Accuracy(EA) - Video targeted edit}{
You are an expert evaluator for audio-video editing quality.
Your task is to assess whether a VIDEO edit was performed correctly.
\newline
{[Source video]}: (attached)\newline
{[Edited video]}: (attached)\newline
\newline\newline
{[Source description]}: "{source\_description}"\newline
{[Target description]}: "{target\_prompt}"\newline
\newline\newline
=== EVALUATION CRITERIA (human-verified) ===\newline
Changes intended:
  "{what\_changed}"\newline

Elements that must remain intact:
  "{what\_preserved}"\newline
============================================
\newline
Evaluate the edited video based on the criteria above.
\newline
Scoring rubric (1-5):\newline
5: All intended changes are fully and precisely realized,
   and all preserved elements remain completely intact. \\
4: Intended changes are mostly realized with minor inaccuracies
   (e.g., correct object but slightly off in appearance or position),
   preserved elements are intact.\\
3: Intended changes are partially realized
   (e.g., some changes applied correctly, others missing),
   OR preserved elements show minor unintended damage.\\
2: Intended changes are attempted but mostly incorrect or incomplete,
   OR preserved elements are significantly damaged.\\
1: No meaningful change was made,
   OR changes go in the wrong direction,
   OR preserved elements are destroyed.\\

Provide your response in this format:\newline
(1) Change assessment: For each intended change, describe whether and how well it was achieved.\\
(2) Preservation assessment: For each preserved element, describe whether it remains intact.\\
(3) Score: [1-5]

}
\promptbox{Edit Accuracy(EA) - Audio-Video coupled edit}{

You are an expert evaluator for audio-video editing quality.
Your task is to assess whether an AUDIO-VIDEO coupled edit
was performed correctly.
\newline\newline
{[Source video with audio]}: (attached)\newline
{[Edited video with audio]}: (attached)
\newline\newline
{[Source description]}: "{source\_description}"\newline
{[Target description]}: "{target\_prompt}"
\newline\newline
=== EVALUATION CRITERIA (human-verified) ===\newline
Changes intended:
  "{what\_changed}"
\newline\newline
Elements that must remain intact:
  "{what\_preserved}"\newline
============================================
\newline\newline
Evaluate BOTH the edited video and audio based on the criteria above.
\newline\newline
Scoring rubric (1-5):\newline
5: All intended changes (both audio and video) are fully and precisely realized, preserved elements remain completely intact. \\
4: Intended changes are mostly realized in both modalities with minor inaccuracies, preserved elements are intact. \\
3: Intended changes are partially realized (e.g., one modality edited well but the other incomplete), OR preserved elements show minor unintended damage. \\
2: Intended changes are attempted but mostly incorrect in one or both modalities, OR preserved elements are significantly damaged. \\
1: No meaningful change was made in either modality, OR changes go in the wrong direction, OR preserved elements are destroyed. \\

Provide your response in this format:\newline
(1) Video change assessment: Describe whether intended visual
    changes were achieved.\newline
(2) Audio change assessment: Describe whether intended audio
    changes were achieved.\newline
(3) Preservation assessment: For each preserved element, describe
    whether it remains intact.\newline
(4) Score: [1-5]

}
\subsubsection{Modality Selectivity}
\promptbox{Modality Selectivity (MS) - Audio targeted edit}{You are an expert evaluator for modality preservation
in audio-video editing. \\
This is an AUDIO-ONLY edit. The video must remain UNCHANGED.
Your task is to assess how well the video was preserved.
\\\newline
{[Source video]}: (attached)\newline
{[Edited video]}: (attached)
\\\newline
{[Source description]}: "{source\_description}"\newline
{[Target description]}: "{target\_prompt}"
\\\newline
=== PART A: General Preservation Check === \\
Answer each question with Yes (1) or No (0).\\\newline

Q1: Does the same primary subject from the source video
    still exist in the edited video?\\
Q2: Is the primary subject's action/motion preserved
    identically to the source?\\
Q3: Is the background (location, structure, layout)
    identical to the source?\\
Q4: Is the edited video free from any new blur, artifacts,
    or distortions not present in the source?\\
Q5: Are the overall color tone, lighting, and visual style
    preserved identically to the source?
\\\newline
=== PART B: Sample-Specific Preservation Check === \\
The following elements were verified by human annotators
as elements that MUST be preserved:\\
  "{what\_preserved}" \\

Q6: Are ALL of the specified preserved elements fully maintained
    in the edited video without any alteration? (Yes=1 / No=0)
    If No, explain which specific elements were altered and how. \\

============================================ \\

Provide your response in this format:\\
Q1: (brief evidence) - [Yes/No]\newline
Q2: (brief evidence) - [Yes/No]\newline
Q3: (brief evidence) - [Yes/No]\newline
Q4: (brief evidence) - [Yes/No]\newline
Q5: (brief evidence) - [Yes/No] \newline
Q6: (brief evidence, reference specific elements from preservation criteria) - [Yes/No]\\

Total score: [sum] / 6
}
\newpage
\promptbox{Modality Selectivity (MS) - Video targeted edit}{You are an expert evaluator for modality preservation
in audio-video editing. This is a VIDEO-ONLY edit. The audio must remain UNCHANGED.
Your task is to assess how well the audio was preserved.
\\\newline
[Source audio]: (attached)\newline
[Edited audio]: (attached)
\\\newline
[Source description]: "{source\_description}"\newline
[Target description]: "{target\_prompt}"
\\\newline
=== PART A: General Preservation Check ===\newline
Answer each question with Yes (1) or No (0).\newline
\newline
Q1: Does the same primary foreground sound from the source
    still exist in the edited audio?\newline
Q2: Is the background sound (ambient/environmental sound)
    identical to the source?\newline
Q3: Is the volume level maintained similarly to the source?\newline
Q4: Are the timing and rhythm of the sound identical
    to the source?\newline
Q5: Is the edited audio free from any new sounds
    not present in the source?
\\\newline
=== PART B: Sample-Specific Preservation Check ===\newline
The following elements were verified by human annotators
as elements that MUST be preserved:
  "{what\_preserved}"
\\\newline
Q6: Are ALL of the specified preserved elements fully maintained
    in the edited audio without any alteration? (Yes=1 / No=0)
    If No, explain which specific elements were altered and how.
\\
============================================

Provide your response in this format:\newline
Q1: (brief evidence) - [Yes/No]\newline
Q2: (brief evidence) - [Yes/No]\newline
Q3: (brief evidence) - [Yes/No]\newline
Q4: (brief evidence) - [Yes/No]\newline
Q5: (brief evidence) - [Yes/No]\newline
Q6: (brief evidence, reference specific elements from preservation criteria)- [Yes/No]\newline

Total score: [sum] / 6
}
\subsubsection{Perceptual Quality}
\promptbox{Perceptual Quality (PQ) - Audio targeted edit}{You are an expert evaluator for audio quality assessment.
Your task is to assess the perceptual quality of an edited audio track.\\
Do NOT evaluate whether the edit content is correct —
focus ONLY on technical and perceptual quality.
\newline
{[Edited audio]}: (attached) \newline

Assess the following aspects:\\
- Clarity and crispness of the audio\\
- Smoothness of transitions (no abrupt cuts or jumps)\\
- Natural timbre and realistic sound texture\\
- Temporal continuity (no sudden gaps, loops, or repetitions)\\
- Absence of audio artifacts (clipping, noise, distortion,
  echo, robotic sound, unnatural reverb)
\newline\newline
Scoring rubric (1-5):\\
5: Fully natural and artifact-free.
   Sounds like a professional real-world recording. \\
4: Mostly natural with minor artifacts
   (e.g., slight background hiss, barely noticeable clipping). \\
3: Noticeable artifacts but content is clearly recognizable
   (e.g., intermittent noise, partial distortion, 
    slightly unnatural transitions). \\
2: Severe artifacts that significantly degrade listening experience
   (e.g., persistent noise, heavy distortion, robotic timbre,
    repeated audio glitches). \\
1: Nearly unrecognizable or completely broken
   (e.g., extreme static, fully corrupted audio,
    incomprehensible sound). 
\newline\newline
Provide your response in this format: \\
(1) Quality analysis: Describe the audio's quality,
    noting specific strengths and artifacts found. \\
(2) Score: [1-5]}
\promptbox{Perceptual Quality (PQ) - Video targeted edit}{
You are an expert evaluator for video quality assessment.
Your task is to assess the perceptual quality of an edited video.
Do NOT evaluate whether the edit content is correct ---
focus ONLY on technical and perceptual quality.
\newline
{[Edited video]}: (attached)
\newline
Assess the following aspects:\newline
- Clarity and sharpness of the video\newline
- Temporal consistency (absence of flickering between frames)\newline
- Natural appearance of objects and scenes\newline
- Physical plausibility (no impossible geometry, lighting, etc.)\newline
- Absence of visual artifacts (blur, distortion, ghosting,
  color banding, edge artifacts)\newline
\newline
Scoring rubric (1-5):\newline
5: Fully natural and artifact-free.
   Looks like professionally captured real footage.\newline
4: Mostly natural with minor artifacts
   (e.g., slight blur in one area, barely noticeable flickering).\newline
3: Noticeable artifacts but content is clearly recognizable
   (e.g., intermittent flickering, partial distortion,
    unnatural texture in some regions).\newline
2: Severe artifacts that significantly degrade viewing experience
   (e.g., persistent flickering, heavy distortion,
    unnatural object shapes, color corruption).\newline
1: Nearly unrecognizable or completely broken
   (e.g., extreme noise, fully corrupted frames,
    incomprehensible content).\newline
\newline
Provide your response in this format:\newline
(1) Quality analysis: Describe the video's visual quality,
    noting specific strengths and artifacts found.\newline
(2) Score: {[1-5]}
}
\promptbox{Perceptual Quality (PQ) - Audio-Video coupled edit}{

You are an expert evaluator for audio-video quality assessment.
Your task is to assess the perceptual quality of an edited video with audio.
Do NOT evaluate whether the edit content is correct —
focus ONLY on technical and perceptual quality of both modalities. \\
\\\newline
{[Edited video with audio]}: (attached)
\\\newline
Assess the following aspects:\\

Video:\newline
- Clarity and sharpness of the video\newline
- Temporal consistency (absence of flickering between frames)\newline
- Natural appearance of objects and scenes\newline
- Absence of visual artifacts (blur, distortion, ghosting,
  color banding, edge artifacts)\newline
\\\newline
Audio:\newline
- Clarity and crispness of the audio\newline
- Natural timbre and realistic sound texture\newline
- Absence of audio artifacts (clipping, noise, distortion,
  echo, robotic sound, unnatural reverb)\newline
\\
Scoring rubric (1-5):\newline
5: Both video and audio are fully natural and artifact-free.\newline
4: Minor artifacts in one or both modalities,
   does not significantly affect the overall experience.\newline
3: Noticeable artifacts in at least one modality,
   content is still clearly recognizable.\newline
2: Severe artifacts in one or both modalities
   that significantly degrade the experience.\newline
1: One or both modalities are nearly unrecognizable or broken.\newline
\newline
Provide your response in this format:\newline
(1) Video quality analysis: Describe visual quality and artifacts found.\newline
(2) Audio quality analysis: Describe audio quality and artifacts found.\newline
(3) Score: [1-5]

}
\subsubsection{Audio-Video Consistency}
\promptbox{AV Consistency (AV-C)}{You are an expert evaluator for audio-video coherence assessment.
Your task is to assess whether the audio and video in an edited
result are semantically coherent and well-synchronized.
\newline
[Edited video with audio]: (attached)
\\\newline
=== EDIT CONTEXT (human-verified) === \\
The following changes were intended in this AV-coupled edit:
  "{what\_changed}"
===================================== \\\\

Assess the following aspects:\\
- Semantic coherence: Do the audio and video content match
  each other meaningfully? (e.g., if a dog is shown,
  is a dog-related sound heard?)\\
- Temporal synchronization: Are audio events aligned with
  visual events in timing? (e.g., impact sound at the moment
  of visual contact)\\
- Contextual consistency: Given the intended edit, do both
  modalities reflect the changes in a unified, believable way?\\

Scoring rubric (1-5):\\
5: Perfect coherence. Audio and video correspond precisely —
   sounds match visual content, timing is synchronized,\\
   and the intended edit is reflected coherently in both modalities.\newline
4: Mostly coherent with minor misalignment
   (e.g., slight timing offset, or one subtle element
    doesn't quite match between modalities).\\
3: Related but noticeable mismatch
   (e.g., generally correct category of sound for the scene,
    but specific details don't align, or timing is clearly off).\\
2: Weak relation between audio and video
   (e.g., both modalities seem edited independently
    without awareness of each other).\\
1: Completely unrelated or contradictory
   (e.g., visual shows one scene while audio represents
    an entirely different context).\\

Provide your response in this format: \\
(1) Semantic analysis: Describe what is seen in the video
    and what is heard in the audio. \\
(2) Coherence assessment: Analyze how well the audio and video
    correspond to each other, with specific examples. \\
(3) Synchronization assessment: Comment on temporal alignment
    between audio and visual events. \\
(4) Score: [1-5]
}

\end{document}